\documentclass[]{aa}
\usepackage{graphicx}
\usepackage{longtable}
\usepackage{color}
\usepackage{natbib}

\begin{document}

\title{Gyrochronological dating of the stellar moving group \\ Group X }
\author{S.\,Messina\inst{1}
\and 
D.\,Nardiello\inst{2,3}
\and 
S. Desidera \inst{3}
\and
M.\,Baratella\inst{4,3}
\and
S. Benatti\inst{5}
\and
K. Biazzo\inst{6}
\and
V. D'Orazi\inst{3}
}
\offprints{Sergio Messina}
\institute{INAF-Catania Astrophysical Observatory, via S.Sofia, 78 I-95123 Catania, Italy \\
\email{sergio.messina@inaf.it}
\and
Aix Marseille Univ, CNRS, CNES, LAM, Marseille, France 
\and
Istituto Nazionale di Astrofisica - Osservatorio Astronomico di Padova, Vicolo dell'Osservatorio 5, IT-35122, Padova, Italy 
\and
Dipartimento di Fisica e Astronomia {\it Galileo Galilei}, Vicolo dell'Osservatorio 3, I-35122, Padova, Italy 
\and
INAF - Astronomical Observatory of Palermo, Piazza del Parlamento 1, 90134, Palermo, Italy.
\and
INAF - Osservatorio Astronomico di Roma, Via Frascati 33, I-00040, Monte Porzio Catone (RM), Italy \\
}

\date{}
\titlerunning{Group X}
\authorrunning{S.\,Messina et al.}
\abstract {Gyrochronology is one of the methods currently used to  estimate  the age of stellar open clusters. Hundreds of new clusters, associations, and moving groups unveiled by Gaia and complemented by accurate rotation period measurements provided by recent space missions such as Kepler and TESS are allowing us to significantly improve the reliability of this method. }{We use gyrochronology, that is, the calibrated age-mass-rotation relation valid for low-mass stars, to measure the age of the recently discovered  moving group  Group X.}{We extracted the light curves of all candidate members from the TESS full frame images and measured their rotation periods using different period search methods. }{We measured the rotation period of 168 of a total of 218 stars and compared their period-colour distribution with those of two age-benchmark clusters, the Pleiades (125\,Myr) and Praesepe (625\,Myr),  as well as with the recently characterised open cluster NGC\,3532 (300\,Myr). }{As result of our analysis, we derived a gyro age of 300$\pm$60 Myr. We also applied as independent methods the fitting of the entire isochrone and of the three brightest candidate members individually with the most precise stellar parameters, deriving comparable values of 250\,Myr and 290\,Myr, respectively. Our dating of Group X allows us to definitively rule out the previously proposed connection with the nearby but much older Coma Berenices cluster.}
\maketitle
\rm

\section{Introduction}

Stellar age is a key parameter in several astrophysical contexts, from exo-planetary science, where the derived values of the planet's physical parameters depend on the age of the host star (see e.g. \citealt{Carleo21}), to Milky Way studies, where Galactic formation and evolution models can be constrained if the age of  numerous field and cluster stars is known (see e.g. \citealt{Hayden20}).
An accurate estimate of the ages of coeval stars in stellar clusters and associations is more reliable if compared to the results obtained for Galactic field stars. Some techniques for the estimation of the ages of cluster members are based on the comparison between measurable stellar parameters and stellar evolutionary models (e.g.  main-sequence and turn-off isochrone fitting; \citealt{Pont04}), and lithium-depletion boundary fitting (\citealt{Stauffer98}, \citealt{Messina16}). Other methods make use of calibrated  empirical relationships (e.g. the gyrochronology; \citealt{Angus19}, \citealt{Barnes07}), specific element abundance  ratios (e.g. \citealt{Maldonado15}), and activity proxies (e.g. \citealt{Zhang19}, \citealt{Messina21}).
Asteroseismic analysis allows us to obtain the age of single stars in our Galaxy (\citealt{Lebreton09}). Firm calibrations are required in the case of methods based on empirical calibrations, and, usually, different approaches are suitable for limited regions of the parameter space,  making age determination a particularly challenging task (see \citealt{Soderblom10} for a review).
However, a combination of different techniques for the estimate of the ages allows us to obtain a final robust result (see e.g. \citealt{Desidera15}).

 The recent Gaia Early Third Data Release (\citealt{Bailer-Jones21}) is unveiling a plethora of stellar open clusters and associations (e.g. \citealt{Cantat-Gaudin18}). Complementary measurements of rotation periods of the candidate members of newly discovered clusters and associations from all-sky ground-based projects (e.g. SuperWASP; \citealt{Pollacco06}) and space-borne missions (e.g. Kepler/K2, \citealt{Borucki18}), not only allow the membership to be solidified through gyrochronology but also provide the opportunity to get a robust calibration of the gyrochronology over a large range of ages.

Usually, as first step, the rotation period-colour distribution of newly discovered clusters is compared with known age-benchmark clusters, such as the Pleiades and Praesepe. Then, the isochrone fitting and any other available age diagnostics are also used to secure consistent results. In a following step, the inferred ages of the new clusters are used to improve the age sampling of the gyrochronology.
 This approach allowed  \citet{Curtis19} to discover the Pisces-Eridanus stellar stream and to estimate  a gyrochronological age of about 120\,Myr and  allowed \citet{Bouma21} to discover a halo for the open cluster NGC\,2516 and estimate a gyrochronological age of about 150\,Myr, to mention just a couple recent studies.
In this framework, we present the results of our analysis of the newly discovered moving group Group X. 

Group X is a nearby moving group (d $\sim$ 101\,pc; \citealt{Tang18}).  Based on the Gaia Data Release 1 TGAS  (Tycho-Gaia Astrometric Solution) data, \citet{Oh17}  first discovered an initial sample of 27 candidate members, which were subsequently confirmed as a group by \citet{Faherty18}. The most recent analysis was carried out by \citet{Tang19}, who discovered up to 218 candidate members,  including the 27 candidates listed by \citet{Oh17}. Moreover, they ruled out the previously proposed connection with the nearby Coma Berenices group, providing clear evidence that they  are two dynamically distinct systems. Group X is an interesting example of a moving group at the final stage of disruption by the Galactic tides, as evidenced by the irregular and elongated space distribution of its members. The isochrone fitting method applied by \citet{Tang19} yielded an age estimate of 400\,Myr.

In this work we used astrometric and kinematic data made available by Gaia, rotation periods from TESS (Transiting Exoplanet Survey Satellite), colour-magnitude diagrams (CMDs) suitable for isochrone fitting, and other age diagnostics (such as the lithium line and  activity indicators), which we are collecting for a selection of members  to estimate the age and to add a new age tick mark in the rotation-mass-age relation of low-mass stars.

In Sect.\,2 we describe the photometric data on which our analysis is based. In Sect. 3 we present the CMD and the results of our period search analysis.
Discussion and conclusions are presented in Sects. 4 and 5.


\section{Data}

We used the data collected by TESS in the second year of its main mission (Sectors~14 - 26) between July 18, 2019, and July 4, 2020.  We obtained the light curves of the stars from the full frame images (FFIs) by using the PATHOS pipeline described in \cite{Nardiello2019}. Briefly, we used the software \texttt{img2lc} (written in \texttt{FORTRAN 90/95} +  \texttt{OPENMP})  developed by \cite{Nardiello2015,Nardiello2016} for ground-based instruments to extract the light curves from the FFIs. This software takes as input the FFIs, empirical point spread function (PSF) arrays, and an input catalogue, and, after modelling and subtracting the neighbour stars to each target source in the input catalogue, it measures the flux of the target star with four different apertures (1-px, 2-px, 3-px, 4-px aperture) and PSF-fitting photometry. Different apertures work better for stars of different magnitudes, and we selected the best aperture for each target, comparing their mean rms distributions as described in detail by \cite{Nardiello2020b}.  The light curves are then corrected by using the cotrending basis vectors as described in \cite{Nardiello2020a, 2021arXiv210509952N}. As input catalogue we used the list of stars published by \cite{Tang19}, which contains 218 likely members of Group~X. We extracted 770 light curves: only 3 stars are observed in a single sector, 18 stars are observed in 2 sectors, 88 stars are observed in 3 sectors, 89 stars are observed in 4 sectors, 14 stars in 5 sectors, 5 stars in 6 sectors, and 1 star is observed in 11 sectors.

Light curves will be released on the Mikulski Archive for Space
Telescopes (MAST) as a High Level Science Product (HLSP) under
the project PATHOS\footnote{https://archive.stsci.edu/hlsp/pathos/}.
A detailed
description of the light curves is given in \cite{Nardiello2019}.

\section{Analysis}

\begin{figure}
\begin{minipage}{10cm}
\includegraphics[scale = 0.38, trim = -40 0 40 100, clip, angle=90]{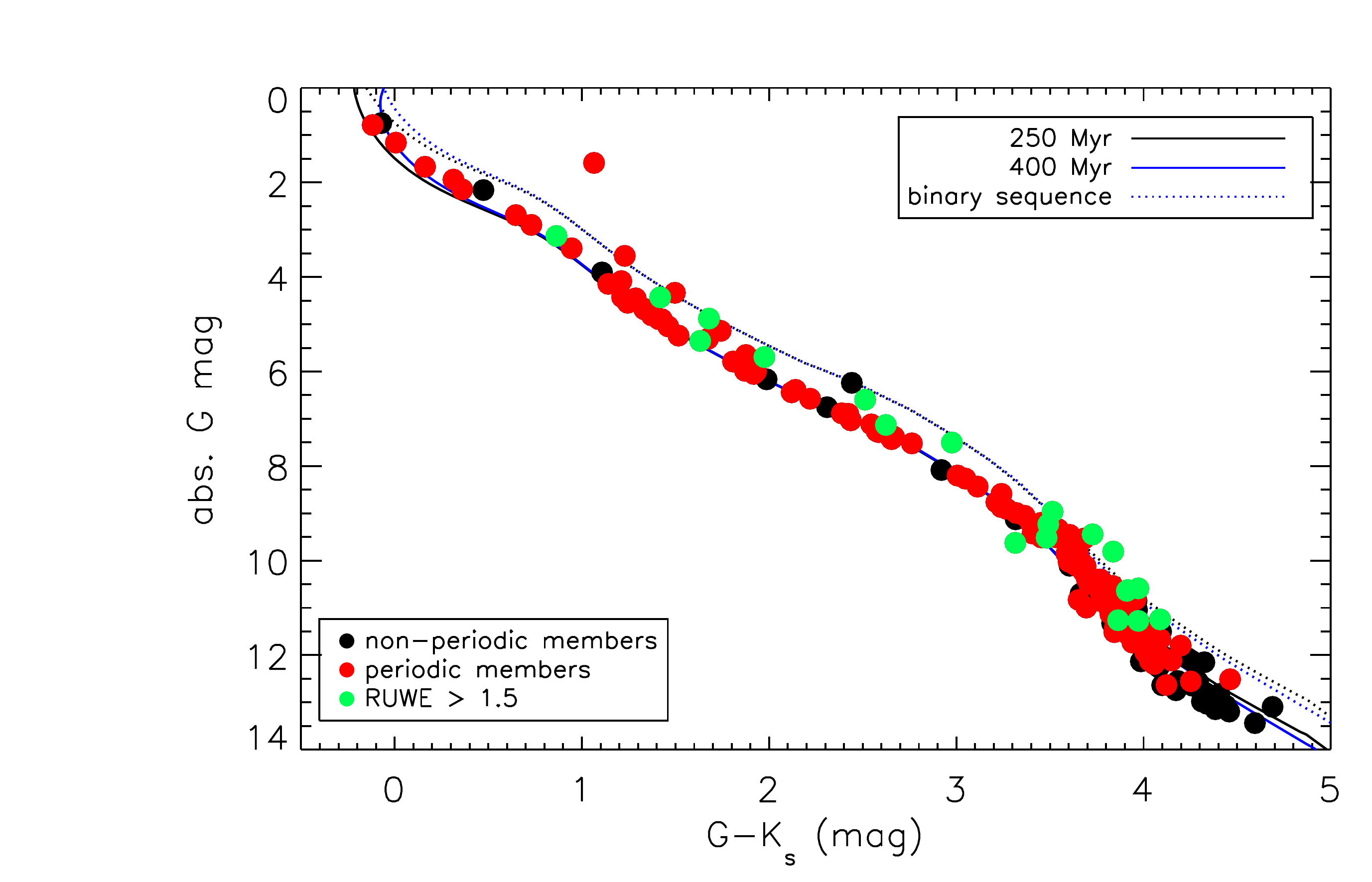}\\
\end{minipage}
\vspace{-1cm}
\caption{\label{hr} CMD for the Group X candidate members (\citealt{Tang19}), with the 250 Myr (solid black line) and 400 Myr (solid blue line) isochrones overplotted. Dashed lines represent the sequence of equal-mass main-sequence binaries. Errors on magnitudes and colours are  smaller than the symbol size. }
\end{figure}

\subsection{Colour-magnitude diagram}
The CMD presented in Fig.\,\ref{hr} was obtained using the Gaia Data Release 2 (DR2) parallaxes and G magnitudes complemented with 2MASS (Two Micron All-Sky Survey) K$_s$ magnitudes (\citealt{Cutri03}). Black and red bullets are used to distinguish between non-periodic and periodic candidate members, respectively (see the following subsection). Seven periodic candidate members with a nearby companion (separation $\rho$ $<$ 3$^{\prime\prime}$) are unresolved in the 2MASS photometry. For these candidates, we computed the correction to be applied to the G$-$K$_s$ colour, using the G magnitudes and parallaxes  of the components and  the PARSEC models of \citet{Bressan12} computed for an age of 400\,Myr and deriving the expected K$_s$ magnitudes for both components (see e.g. Sect.\,3 in \citealt{Messina19}). Finally, colours were corrected for interstellar reddening. We first calculated the E(B$-$V) of each star 
by using the \texttt{PYTHON} routine \texttt{mwdust}\footnote{https://github.com/jobovy/mwdust} (\citealt{Bovy2016}) and the
Combined19 dust map (\citealt{Drimmel2003,Marshall2006,Green2019}), and then we transformed it into E(G$-$K) according to the method presented in \cite{Bessell88}.
 We found that the correction for the interstellar reddening results in a slightly smaller scatter from the isochrone (see ahead) if the average value, $<$E(G$-$K$_s$)$>$ = 0.015$\pm$0.006\,mag (E(B$-$V) = 0.005\,mag), is applied instead of correcting each target for its own reddening. For instance, this reddening is very close to the null reddening computed for the nearby Coma Berenices cluster \citep{Tang18}.
In the following analysis, the average reddening is applied.

The CMD was compared with a series of isochrones based on PARSEC models of \citet{Bressan12}, which span a range of ages from 150\,Myr to 700\,Myr.
The isochrone that best describes the observed CMD has an age of 250 Myr. A slightly better description, but of only the upper bluer part of the CMD, is provided by the isochrone of 400\,Myr.

We also plot, as dotted lines, the sequences of equal-mass binaries for both ages.
We note a number of candidates (27) that lie either significantly above the equal-mass binary isochrone (\#89, \#102, and \#103) or in the magnitude interval between the single and the equal-mass binary sequence. To further investigate their nature and to unveil the presence of unresolved close binaries among them,
we used the re-normalised unit weight error (RUWE; see \citealt{Lindegren18}). All candidates with RUWE $> 1.5$ (a total of 22) are overplotted in green. The following candidates, although significantly displaced from the single star sequence, were not classified according to the RUWE as close binaries: \#27, \#38, \#40, \#47, \#58, \#102, \#143, \#158, \#170, \#173, \#183, \#184, \#201, and \#208.

 These outliers may be still members but may be suffering from an underestimated correction for reddening. Justifying the position in the CMD of star \#102 is more problematic.
On the other side, the inability of the isochrone to adequately fit the bottom end of the sequence is a known problem (see e.g. \citealt{Bell12}, \citealt{Morrell19}), and its discussion is beyond the scope of the present study.

\begin{figure*}
\begin{minipage}{10cm}
\centering
\includegraphics[scale = 0.50, trim = 0 0 0 0, clip, angle=90]{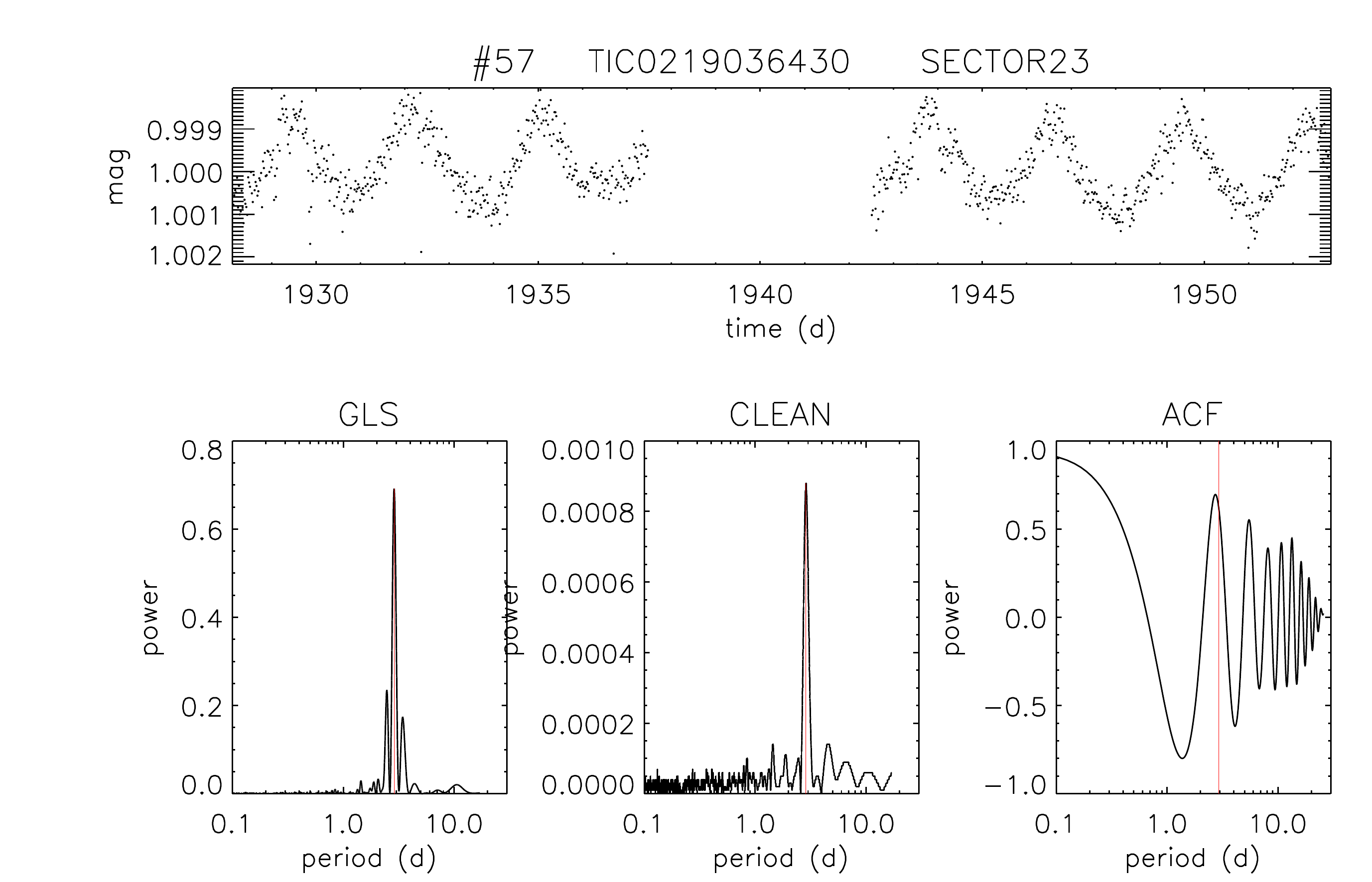}\\
\end{minipage}
\vspace{1cm}
\caption{\label{lightcurve} Example case of periodogram analysis of target \#57. Top panel: TESS magnitude time series. Bottom: GLS periodogram (left), CLEAN periodogram (middle), and ACF (right), with the solid vertical  red  lines indicating the rotation period.}
\end{figure*}
As a complementary approach to estimating the age of Group X, we also considered the three brightest candidate members individually with the most precise determination of their stellar parameters (\#51: 84 UMa, \#28: HD 118214, and \#45: HD 119765), exploiting the PARAM online tool for the Bayesian estimation of stellar parameters \citep{dasilva2006} 
to obtain the most probable age.
The individual ages are 240$\pm$100, 320$\pm$110, and 300$\pm$110 Myr, respectively.
This supports an age with a mean value of 290\,Myr as the most probable one.

\subsection{Rotation period search}

\begin{figure*}
\begin{minipage}{10cm}
\includegraphics[scale = 0.50, trim = 0 0 0 0, clip, angle=90]{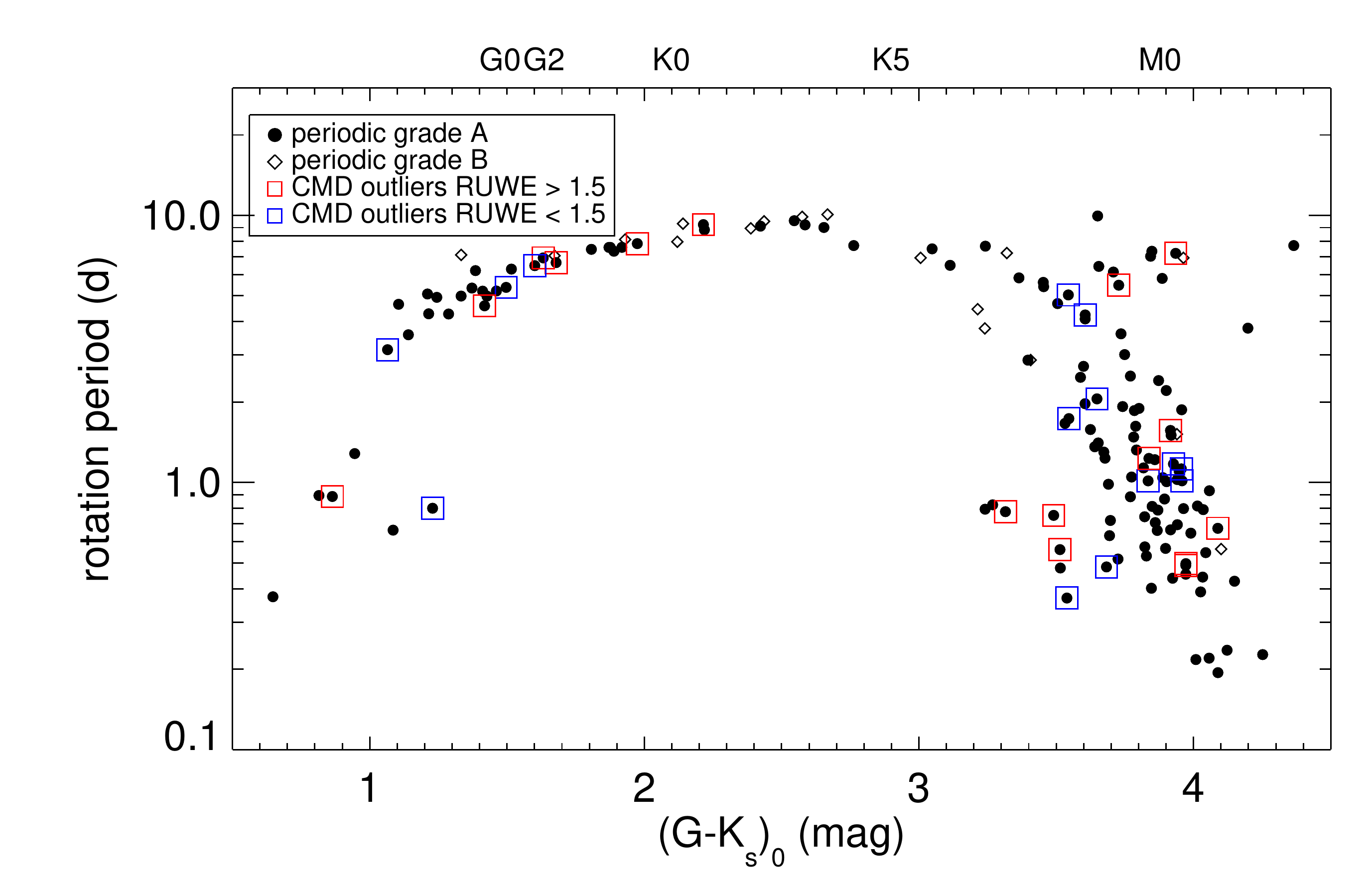}\\
\end{minipage}
\caption{\label{color-period} Distribution of stellar rotation periods versus de-reddened colour for the periodic candidate members of Group X.}
\end{figure*}

We analysed the TESS light curves of the Group X candidate members to measure the rotation period using three different methods: generalised Lomb-Scargle (GLS; \citealt{Zechmeister09}), CLEAN (\citealt{Roberts87}), and the autocorrelation function (ACF; \citealt{Mcquillan13}). Details and examples of the use of these methods can be found in \cite{Messina17} for GLS and CLEAN, and \cite{Mcquillan13} for ACF. We used more than one method in order to provide a `grade' of confidence on the correctness of the measured rotation periods.
If the values of the rotation periods were found by all three methods to be similar  within the respective uncertainties, we assigned a quality grade `A'; when only two methods found the same value, we assigned a grade `B'. Period estimates differing in all three methods were not considered. Since most stars were observed in more TESS sectors, our period search was performed in each available sector. Generally, the same rotation period was found in subsequent sectors,  increasing the robustness of our measurement results. An example case of periodogram analysis is given in Fig.\,\ref{lightcurve}.

We selected only rotation periods with false alarm probability (FAP) $<$ 0.1\%. The FAP was computed using the analytical formulae of \cite{Horne86}, which are valid for evenly spaced time series data. We followed the
method used by \cite{Lamm04}  to compute the errors associated with the period determinations (see e.g. \citealt{Messina10}, for details). 
From a total sample of 218 stars, we measured 150 periods with grade A and 18 periods with grade B. In Fig.\,\ref{color-period} we plot the rotation period distribution of Group X. We use different symbols to indicate grade A (filled circles) and grade B (diamonds) periods and  squared symbols to indicate the mentioned  outliers in the CMD (red for RUWE $>$ 1.5 and blue for RUWE $<$ 1.5).  In Table\,\ref{tab-period} we list the rotation periods with respective uncertainty, grade, and the TESS  sector in which the same value of period was measured.

As mentioned, the PATHOS pipeline subtracts from the target the flux of any neighbour star in the adjacent pixels (i.e. at distances $\rho$ $>$ 10$^{\prime\prime}$). As a consequence of this and the dilution effect, any effect of variability in the residual flux of the nearby stars becomes negligible. On the contrary, the flux of neighbour stars at distances $\rho$ $<$ 10$^{\prime\prime}$ is not removed by the pipeline and it may contribute significantly to the observed variability. 

A total of 17 periodic stars in our sample (marked with an asterisk in Table\,\ref{tab-period}) have one visual companion detected in the Gaia DR2 but are unresolved in the TESS photometry at a separation of $\rho$ $\la$ 10$^{\prime\prime}$ and with a magnitude difference of $\Delta$G $<$ 3\,mag (all have RUWE $<$ 1.5). 
 We inspected the periodograms of these 17 candidate members and found nine cases (stars \#12, \#17, \#27, \#137, and \#141 and the systems \#29\&30 and \#148\&149) with a significant secondary period. The secondary period can be interpreted as the rotation period of the nearby companion (see e.g. \citealt{Messina19}, \citealt{Bonavita21}, \citealt{Tokovinin18}). On the other hand, stars \#39 and \#63, which have no detected nearby companion in the Gaia DR2, have clear evidence of a secondary period, P = 0.3639\,d and P = 0.4646\,d, respectively.\\

\rm


\section{Discussion}

As mentioned in Sect.\,1, the main scope of the present work is to estimate the age of Group X via gyrochronology,  providing independent support to the membership by \citet{Tang19}.  This is accomplished by comparing the distribution of the rotation period with those of  primary  age-benchmark open clusters, specifically the Pleiades with an age of $\sim$125\,Myr (\citealt{Stauffer98}) and Praesepe with an age of $\sim$625\,Myr (\citealt{Brandt15}). A comparison with the NGC\,3532 ($\sim$300\,Myr; \citealt{Fritzewski21}) and M48 ($\sim$450\,Myr; \citealt{Barnes15}) clusters is also done. 
 Rotation periods of Pleiades members are taken from \citet{Rebull16}, those of Praesepe from \citet{Rebull17},  those of NGC\,3532 from \citet{Fritzewski21}, and those of M48 from \citet{Barnes15}.

Colours were de-reddened by adopting a colour excess E(B$-$V) = 0.045\,mag  for the Pleiades and E(B$-$V) = 0.027\,mag for Praesepe (\citealt{Gaia_Collaboration18}) and   E(B$-$V) = 0.035\,mag for NGC\,3532 (\citealt{Fritzewski21}) and  E(B$-$V) = 0.08\,mag for M48 (\citealt{Barnes15}). 
Finally, (G$-$K$_s$)$_0$ colours were transformed into B$-$V colours using the calibration by \citet{Pecaut13}.

\begin{figure*}
\begin{minipage}{10cm}
\centering
\includegraphics[scale = 0.50, trim = 0 0 0 0, clip, angle=90]{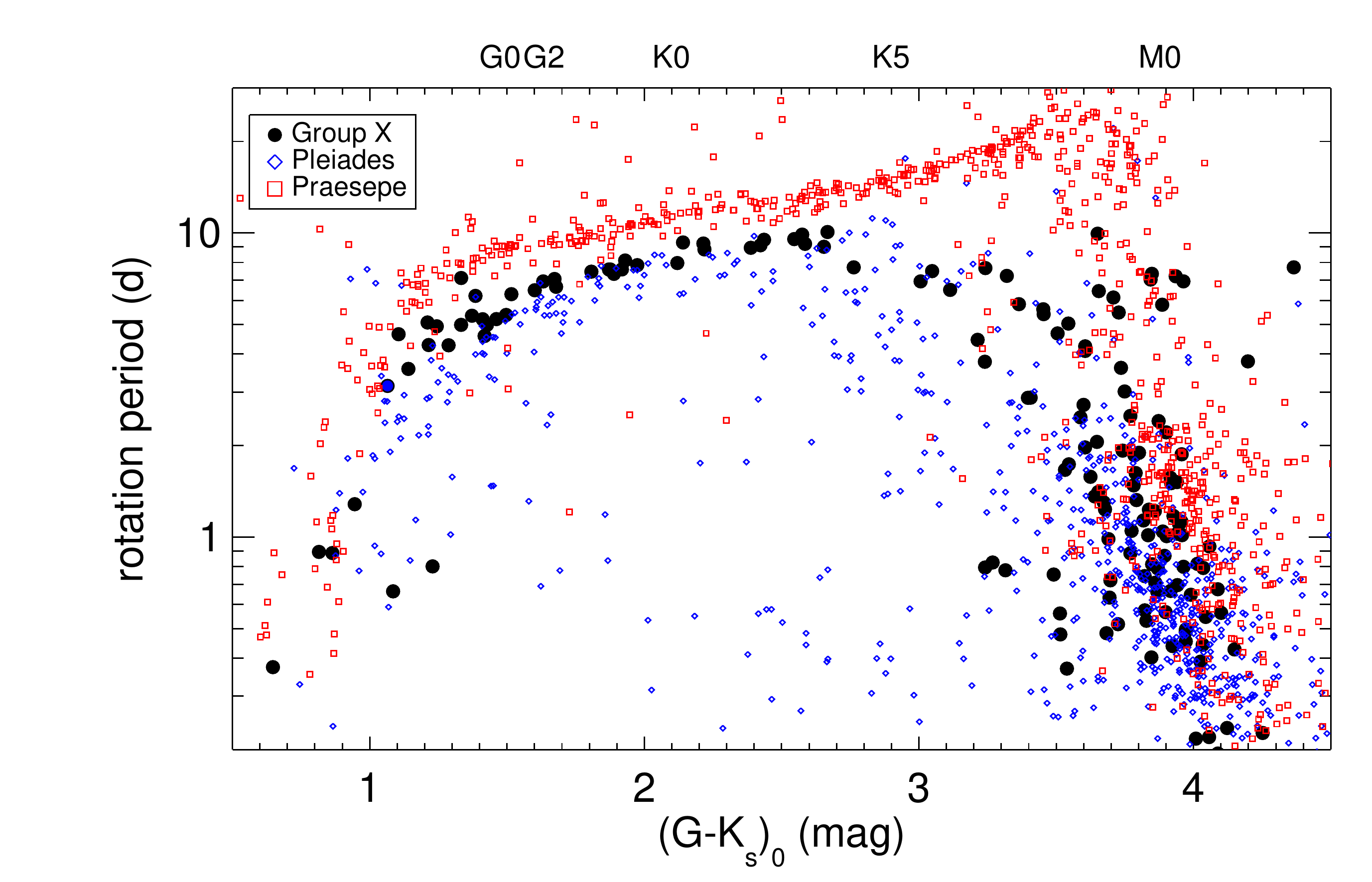}\\
\includegraphics[scale = 0.50, trim = 0 0 0 0, clip, angle=90]{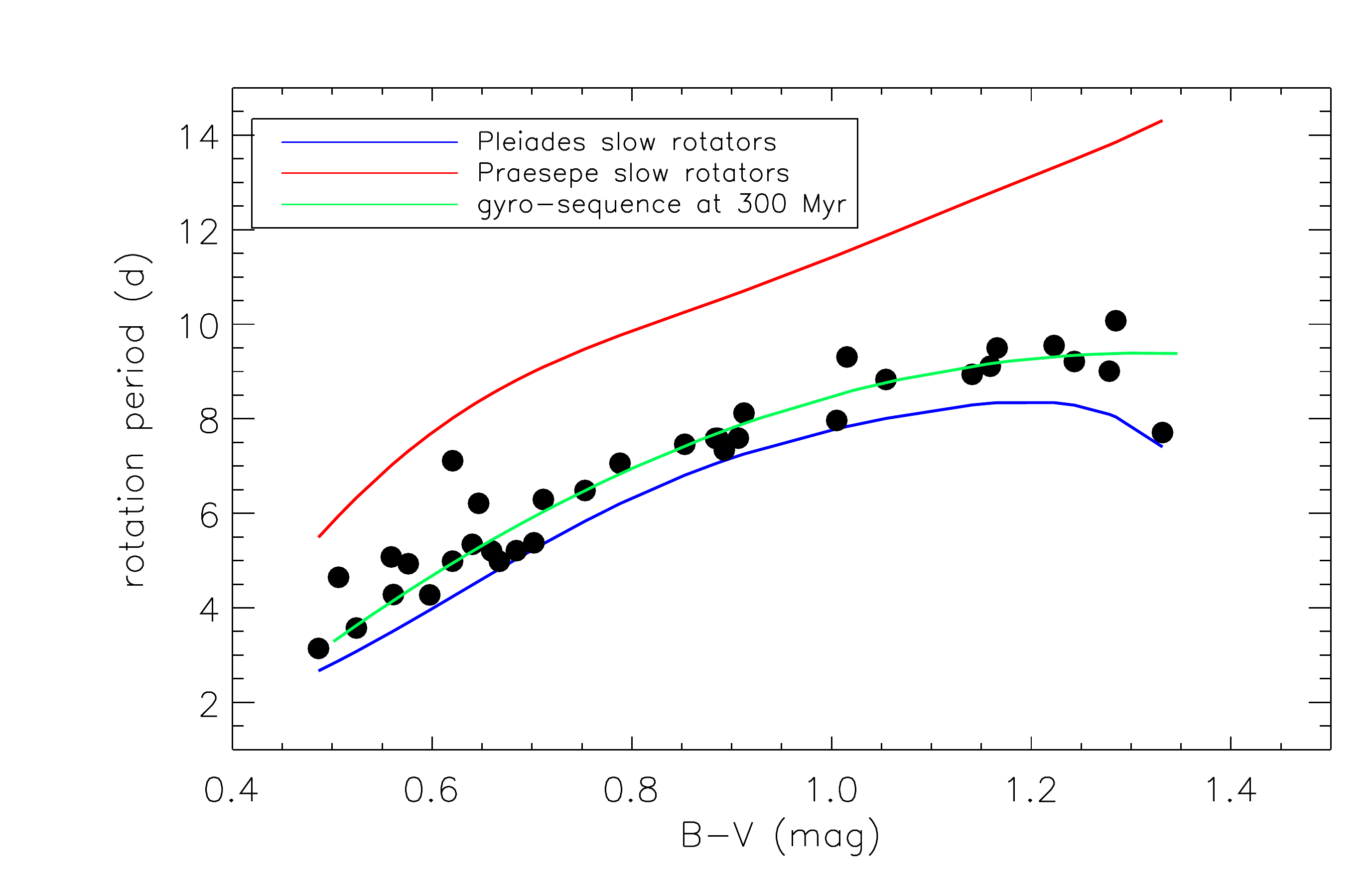}\\
\end{minipage}
\vspace{1cm}
\caption{\label{color-period2} Comparison of colour-period distribution of Group X with age-benchmark open clusters. Top panel: Distribution of stellar rotation periods of Group X with the Pleiades and Praesepe distributions overplotted. Bottom panel: Distribution of the rotation periods of slow rotators of Group X versus B$-$V colour, with the polynomial fits to  median rotation periods of slow rotators in the Praesepe (red line) and Pleiades (blue line) clusters and, as an example, the gyro-sequence corresponding to an age of 300\,Myr (green line), according to the \citet{Mamajek08} coefficients, overplotted.}
\end{figure*}

 To compare the period distribution of Group X with those of the Pleiades and Praesepe  and to derive a quantitative estimate of the age of Group X, we selected the sequence of slow rotators, that is, the colour range 0.5 $<$ (B$-$V)$_0$ $<$ 1.3\,mag, where the dependence on the age of the rotation period is better defined (an almost one-to-one correspondence between colour and period).  As shown in Fig.\,\ref{color-period}, both single stars and candidate binaries that lie between the single and binary sequences follow the same period distribution. Therefore, we opted to include all of them in the process of age estimate.  We adopted a colour binning of 0.10 mag, then for each bin we computed the median rotation period, and, finally, we fitted  a polynomial to the sequence of median values (the solid blue and red lines in the bottom panel of Fig.\,\ref{color-period2}). 
We assumed that in the age interval between the Pleiades and Praesepe the period slowdown has a functional form of the type
\begin{equation}
    P = A^n \times a(B-V-c)^b
.\end{equation}

\begin{figure*}
\begin{minipage}{10cm}
\centering
\includegraphics[scale = 0.50, trim = 0 0 0 0, clip, angle=90]{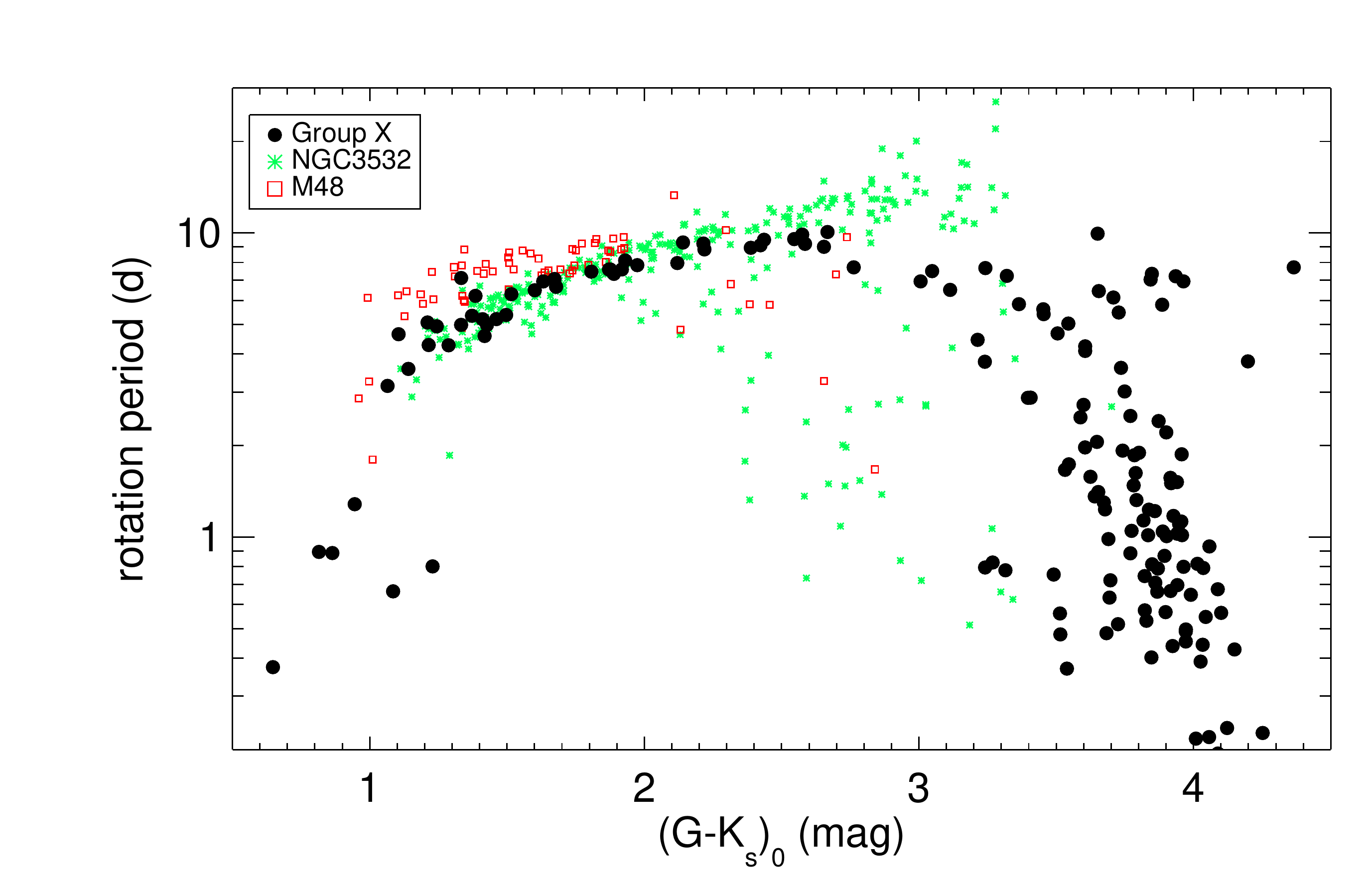}\\
\end{minipage}
\vspace{1cm}
\caption{\label{color-period3} Comparison of the rotation period distributions of Group X with the coeval NGC\,3532 and with the older M\,48 open clusters.}
\end{figure*}

\begin{figure*}
\begin{minipage}{10cm}
\centering
\includegraphics[scale = 0.50, trim = 0 0 0 0, clip, angle=90]{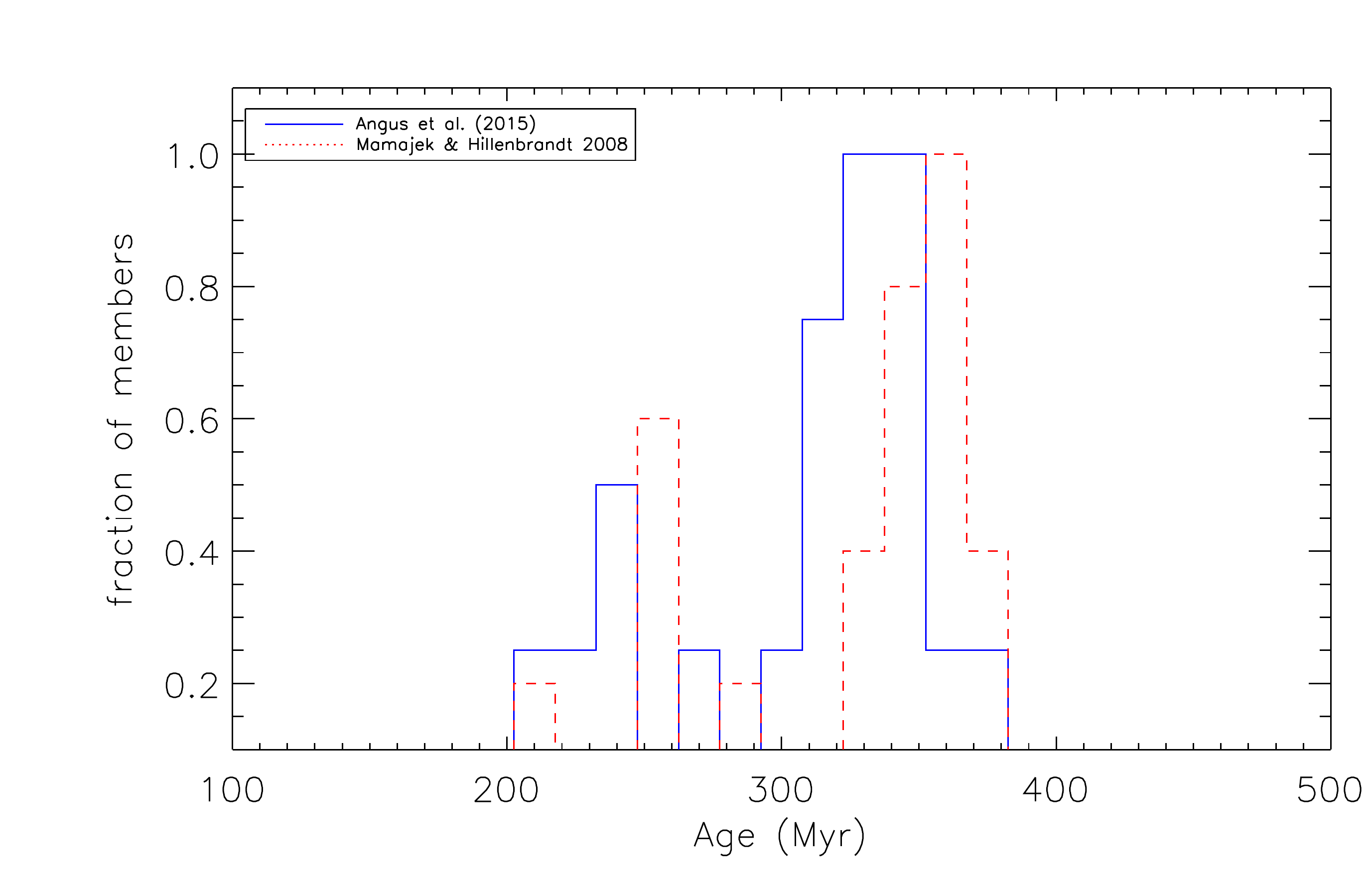}
\end{minipage}
\vspace{1cm}
\caption{\label{distri_age} Distribution of ages of individual candidate members as derived by the relations of \citet{Angus15} (solid blue  line) and by \citet{Mamajek08} (dotted red line).}
\end{figure*}

We inferred the age of each candidate member, finding, as shown in Fig.\,\ref{distri_age},  a bimodal-like distribution with the bulk of members at an age of 350\,Myr ($\sim$360\,Myr using \citet{Mamajek08} coefficients) and a minority, $\sim$25\%, at an age of about 230\,Myr.
For the whole moving group, we obtain an average value of 309$\pm$60\,Myr using the a, b, c, and n coefficients from \citet{Mamajek08} and 297$\pm$50\,Myr using the \citet{Angus15} coefficients. 
 Their average value of 303$\pm$60\,Myr is in agreement within the uncertainties with the isochronal age derived by us (i.e.  250\,Myr) and in agreement with the age of the three brightest stars derived with PARAM (i.e. 290\,Myr). We note that in the selected colour range for the gyrochronological estimate of age  (0.5 $<$ (B$-$V)$_0$ $<$ 1.3\,mag), \rm all visual binaries have their components at distances $\rho$ $>$ 150\,au, which is sufficiently distant to neglect any effects of tides on the rotation period evolution (see e.g. \citealt{Messina19}).

 Finally, the age of $\sim$300 Myr for Group X is also supported by the comparison with the rotation period distribution of the NGC\,3532 open cluster, which has an estimated age of 300\,Myr. As shown in Fig.\,\ref{color-period3}, the two distributions are almost undistinguishable for (G$-$K$_s$)$_0$ $<$ 2.3 mag.  However, in the colour range 2.3 $<$ (G$-$K$_s$)$_0$ $<$ 3.3 mag,
the Group X candidate members, which are all found to be periodic, all rotate faster than their counterpart slow-rotator members of NGC\,3532.  In the mentioned colour range, all candidate members  being periodic, it is unlikely that the absence of longer rotation periods arises from the insensitivity of TESS data to rotation periods longer than 10-12 days. 
\rm

It is worth noting that the possibility of measuring the age of stellar clusters by means of gyrochronology makes the search for exo-planets around their members especially relevant.  This is the case of one TESS target of interest  identified among the candidate members of Group X and whose characterisation (Nardiello et al., in preparation) greatly benefits from the age determined in the present study.

\section{Conclusions}
We have explored the rotational properties of the late-type candidate members of the recently discovered  moving group  Group X, which has a total of 218 candidate members. All the candidate members were observed by TESS in one or more sectors, and we extracted the light curves from the  FFIs using the PATHOS pipeline. The rotation period search was done using three different methods, GLS, CLEAN, and ACF, which provided rotation period measurements for 168 stars (150 with grade A and 18 with grade B). The colour-period distribution was compared with those of two age-benchmark clusters, the Pleiades with a quoted age of 125\,Myr and Praesepe with a quoted age of 625\,Myr. Assuming a temporal evolution of the rotation period in this age range as expressed by Eq. (1), we inferred for Group X an age of 300 $\pm$ 60\,Myr. The comparison was limited to the slow rotators whose rotation period minimum dispersion allows a more accurate comparison among clusters of different ages. Our age estimate is further supported by the similarity of the period distribution with that of NGC\,3532, an open cluster of 300 Myr.  The gyro age we derived is in agreement with the isochronal age of 250 Myr derived by us and definitively younger than the 400\,Myr age previously estimated by \cite{Tang19}. Our dating of Group X allows us to definitively rule out the previously proposed connection with the nearby Coma Berenices cluster ($\sim$700--800\,Myr), further confirming the earlier conclusions by \cite{Tang19}  that they are two dynamically distinct systems.\\

{\it Acknowledgments}
Research on stellar activity at INAF is supported by MUR (Ministero
dell’Università e della Ricerca). This research has made use of the Simbad
database operated at CDS (Strasbourg, France). Financial support from the INAF-ASI agreement (n.2018-16-HH.0) and the PRIN-INAF "PLATEA" (P.I. S. Desidera) are acknowledged.
 DN acknowledges the support from the French Centre National d'Etudes
Spatiales (CNES). SM thanks the Referee whose comments  helped to improve the quality of the paper.

\bibliographystyle{aa.bst} 
\bibliography{biblio.bib} 

\begin{thebibliography}{59}
\expandafter\ifx\csname natexlab\endcsname\relax\def\natexlab#1{#1}\fi

\bibitem[{{Angus} {et~al.}(2015){Angus}, {Aigrain}, {Foreman-Mackey}, \&
  {McQuillan}}]{Angus15}
{Angus}, R., {Aigrain}, S., {Foreman-Mackey}, D., \& {McQuillan}, A. 2015,
  \mnras, 450, 1787

\bibitem[{{Angus} {et~al.}(2019){Angus}, {Morton}, {Foreman-Mackey}, {van
  Saders}, {Curtis}, {Kane}, {Bedell}, {Kiman}, {Hogg}, \& {Brewer}}]{Angus19}
{Angus}, R., {Morton}, T.~D., {Foreman-Mackey}, D., {et~al.} 2019, \aj, 158,
  173

\bibitem[{{Bailer-Jones} {et~al.}(2021){Bailer-Jones}, {Rybizki}, {Fouesneau},
  {Demleitner}, \& {Andrae}}]{Bailer-Jones21}
{Bailer-Jones}, C.~A.~L., {Rybizki}, J., {Fouesneau}, M., {Demleitner}, M., \&
  {Andrae}, R. 2021, \aj, 161, 147

\bibitem[{{Barnes}(2007)}]{Barnes07}
{Barnes}, S.~A. 2007, \apj, 669, 1167

\bibitem[{{Barnes} {et~al.}(2015){Barnes}, {Weingrill}, {Granzer}, {Spada}, \&
  {Strassmeier}}]{Barnes15}
{Barnes}, S.~A., {Weingrill}, J., {Granzer}, T., {Spada}, F., \& {Strassmeier},
  K.~G. 2015, \aap, 583, A73

\bibitem[{{Bell} {et~al.}(2012){Bell}, {Naylor}, {Mayne}, {Jeffries}, \&
  {Littlefair}}]{Bell12}
{Bell}, C. P.~M., {Naylor}, T., {Mayne}, N.~J., {Jeffries}, R.~D., \&
  {Littlefair}, S.~P. 2012, \mnras, 424, 3178

\bibitem[{{Bessell} \& {Brett}(1988)}]{Bessell88}
{Bessell}, M.~S. \& {Brett}, J.~M. 1988, \pasp, 100, 1134

\bibitem[{{Bonavita} {et~al.}(2021){Bonavita}, {Gratton}, {Desidera},
  {Squicciarini}, {D'Orazi}, {Zurlo}, {Biller}, {Chauvin}, {Fontanive},
  {Janson}, {Messina}, {Menard}, {Meyer}, {Vigan}, {Avenhaus}, {Asensio
  Torres}, {Beuzit}, {Boccaletti}, {Bonnefoy}, {Brandner}, {Cantalloube},
  {Cheetham}, {Cudel}, {Daemgen}, {Delorme}, {Desgrange}, {Dominik}, {Engler},
  {Feautrier}, {Feldt}, {Galicher}, {Garufi}, {Gasparri}, {Ginski}, {Girard},
  {Grandjean}, {Hagelberg}, {Henning}, {Hunziker}, {Kasper}, {Keppler},
  {Lagadec}, {Lagrange}, {Langlois}, {Lannier}, {Lazzoni}, {Le Coroller},
  {Ligi}, {Lombart}, {Maire}, {Mazevet}, {Mesa}, {Mouillet}, {Moutou},
  {Muller}, {Peretti}, {Perrot}, {Petrus}, {Potier}, {Ramos}, {Rickman},
  {Rouan}, {Salter}, {Samland}, {Schmidt}, {Sissa}, {Stolker}, {Szulagyil},
  {Udry}, \& {Wildi}}]{Bonavita21}
{Bonavita}, M., {Gratton}, R., {Desidera}, S., {et~al.} 2021, arXiv e-prints,
  arXiv:2103.13706

\bibitem[{{Borucki}(2018)}]{Borucki18}
{Borucki}, W.~J. 2018, {Space Missions for Exoplanet Science: Kepler/K2}, ed.
  H.~J. {Deeg} \& J.~A. {Belmonte}, 80

\bibitem[{{Bouma} {et~al.}(2021){Bouma}, {Curtis}, {Hartman}, {Winn}, \&
  {Bakos}}]{Bouma21}
{Bouma}, L.~G., {Curtis}, J.~L., {Hartman}, J.~D., {Winn}, J.~N., \& {Bakos},
  G.~{\'A}. 2021, \aj, 162, 197

\bibitem[{{Bovy} {et~al.}(2016){Bovy}, {Rix}, {Green}, {Schlafly}, \&
  {Finkbeiner}}]{Bovy2016}
{Bovy}, J., {Rix}, H.-W., {Green}, G.~M., {Schlafly}, E.~F., \& {Finkbeiner},
  D.~P. 2016, \apj, 818, 130

\bibitem[{{Brandt} \& {Huang}(2015)}]{Brandt15}
{Brandt}, T.~D. \& {Huang}, C.~X. 2015, \apj, 807, 24

\bibitem[{{Bressan} {et~al.}(2012){Bressan}, {Marigo}, {Girardi}, {Salasnich},
  {Dal Cero}, {Rubele}, \& {Nanni}}]{Bressan12}
{Bressan}, A., {Marigo}, P., {Girardi}, L., {et~al.} 2012, \mnras, 427, 127

\bibitem[{{Cantat-Gaudin} {et~al.}(2018){Cantat-Gaudin}, {Jordi}, {Vallenari},
  {Bragaglia}, {Balaguer-N{\'u}{\~n}ez}, {Soubiran}, {Bossini}, {Moitinho},
  {Castro-Ginard}, {Krone-Martins}, {Casamiquela}, {Sordo}, \&
  {Carrera}}]{Cantat-Gaudin18}
{Cantat-Gaudin}, T., {Jordi}, C., {Vallenari}, A., {et~al.} 2018, \aap, 618,
  A93

\bibitem[{{Carleo} {et~al.}(2021){Carleo}, {Desidera}, {Nardiello},
  {Malavolta}, {Lanza}, {Livingston}, {Locci}, {Marzari}, {Messina}, {Turrini},
  {Baratella}, {Borsa}, {D'Orazi}, {Nascimbeni}, {Pinamonti}, {Rainer}, {Alei},
  {Bignamini}, {Gratton}, {Micela}, {Montalto}, {Sozzetti}, {Squicciarini},
  {Affer}, {Benatti}, {Biazzo}, {Bonomo}, {Claudi}, {Cosentino}, {Covino},
  {Damasso}, {Esposito}, {Fiorenzano}, {Frustagli}, {Giacobbe}, {Harutyunyan},
  {Leto}, {Magazz{\`u}}, {Maggio}, {Mainella}, {Maldonado}, {Mallonn},
  {Mancini}, {Molinari}, {Molinaro}, {Pagano}, {Pedani}, {Piotto}, {Poretti},
  {Redfield}, \& {Scandariato}}]{Carleo21}
{Carleo}, I., {Desidera}, S., {Nardiello}, D., {et~al.} 2021, \aap, 645, A71

\bibitem[{{Curtis} {et~al.}(2019){Curtis}, {Ag{\"u}eros}, {Mamajek}, {Wright},
  \& {Cummings}}]{Curtis19}
{Curtis}, J.~L., {Ag{\"u}eros}, M.~A., {Mamajek}, E.~E., {Wright}, J.~T., \&
  {Cummings}, J.~D. 2019, \aj, 158, 77

\bibitem[{{Cutri} {et~al.}(2003){Cutri}, {Skrutskie}, {van Dyk}, {Beichman},
  {Carpenter}, {Chester}, {Cambresy}, {Evans}, {Fowler}, {Gizis}, {Howard},
  {Huchra}, {Jarrett}, {Kopan}, {Kirkpatrick}, {Light}, {Marsh}, {McCallon},
  {Schneider}, {Stiening}, {Sykes}, {Weinberg}, {Wheaton}, {Wheelock}, \&
  {Zacarias}}]{Cutri03}
{Cutri}, R.~M., {Skrutskie}, M.~F., {van Dyk}, S., {et~al.} 2003, VizieR Online
  Data Catalog, II/246

\bibitem[{{da Silva} {et~al.}(2006){da Silva}, {Girardi}, {Pasquini},
  {Setiawan}, {von der L{\"u}he}, {de Medeiros}, {Hatzes}, {D{\"o}llinger}, \&
  {Weiss}}]{dasilva2006}
{da Silva}, L., {Girardi}, L., {Pasquini}, L., {et~al.} 2006, \aap, 458, 609

\bibitem[{{Desidera} {et~al.}(2015){Desidera}, {Covino}, {Messina}, {Carson},
  {Hagelberg}, {Schlieder}, {Biazzo}, {Alcal{\'a}}, {Chauvin}, {Vigan},
  {Beuzit}, {Bonavita}, {Bonnefoy}, {Delorme}, {D'Orazi}, {Esposito}, {Feldt},
  {Girardi}, {Gratton}, {Henning}, {Lagrange}, {Lanzafame}, {Launhardt},
  {Marmier}, {Melo}, {Meyer}, {Mouillet}, {Moutou}, {Segransan}, {Udry}, \&
  {Zaidi}}]{Desidera15}
{Desidera}, S., {Covino}, E., {Messina}, S., {et~al.} 2015, \aap, 573, A126

\bibitem[{{Drimmel} {et~al.}(2003){Drimmel}, {Cabrera-Lavers}, \&
  {L{\'o}pez-Corredoira}}]{Drimmel2003}
{Drimmel}, R., {Cabrera-Lavers}, A., \& {L{\'o}pez-Corredoira}, M. 2003, \aap,
  409, 205

\bibitem[{{Faherty} {et~al.}(2018){Faherty}, {Bochanski}, {Gagn{\'e}},
  {Nelson}, {Coker}, {Smithka}, {Desir}, \& {Vasquez}}]{Faherty18}
{Faherty}, J.~K., {Bochanski}, J.~J., {Gagn{\'e}}, J., {et~al.} 2018, \apj,
  863, 91

\bibitem[{{Fritzewski} {et~al.}(2021){Fritzewski}, {Barnes}, {James}, \&
  {Strassmeier}}]{Fritzewski21}
{Fritzewski}, D.~J., {Barnes}, S.~A., {James}, D.~J., \& {Strassmeier}, K.~G.
  2021, \aap, 652, A60

\bibitem[{{Gaia Collaboration} {et~al.}(2018){Gaia Collaboration}, {Babusiaux},
  {van Leeuwen}, {Barstow}, {Jordi}, {Vallenari}, {Bossini}, {Bressan},
  {Cantat-Gaudin}, {van Leeuwen}, {Brown}, {Prusti}, {de Bruijne},
  {Bailer-Jones}, {Biermann}, {Evans}, {Eyer}, {Jansen}, {Klioner}, {Lammers},
  {Lindegren}, {Luri}, {Mignard}, {Panem}, {Pourbaix}, {Randich}, {Sartoretti},
  {Siddiqui}, {Soubiran}, {Walton}, {Arenou}, {Bastian}, {Cropper}, {Drimmel},
  {Katz}, {Lattanzi}, {Bakker}, {Cacciari}, {Casta{\~n}eda}, {Chaoul}, {Cheek},
  {De Angeli}, {Fabricius}, {Guerra}, {Holl}, {Masana}, {Messineo}, {Mowlavi},
  {Nienartowicz}, {Panuzzo}, {Portell}, {Riello}, {Seabroke}, {Tanga},
  {Th{\'e}venin}, {Gracia-Abril}, {Comoretto}, {Garcia-Reinaldos}, {Teyssier},
  {Altmann}, {Andrae}, {Audard}, {Bellas-Velidis}, {Benson}, {Berthier},
  {Blomme}, {Burgess}, {Busso}, {Carry}, {Cellino}, {Clementini}, {Clotet},
  {Creevey}, {Davidson}, {De Ridder}, {Delchambre}, {Dell'Oro}, {Ducourant},
  {Fern{\'a}ndez-Hern{\'a}ndez}, {Fouesneau}, {Fr{\'e}mat}, {Galluccio},
  {Garc{\'\i}a-Torres}, {Gonz{\'a}lez-N{\'u}{\~n}ez}, {Gonz{\'a}lez-Vidal},
  {Gosset}, {Guy}, {Halbwachs}, {Hambly}, {Harrison}, {Hern{\'a}ndez},
  {Hestroffer}, {Hodgkin}, {Hutton}, {Jasniewicz}, {Jean-Antoine-Piccolo},
  {Jordan}, {Korn}, {Krone-Martins}, {Lanzafame}, {Lebzelter}, {L{\"o}ffler},
  {Manteiga}, {Marrese}, {Mart{\'\i}n-Fleitas}, {Moitinho}, {Mora}, {Muinonen},
  {Osinde}, {Pancino}, {Pauwels}, {Petit}, {Recio-Blanco}, {Richards},
  {Rimoldini}, {Robin}, {Sarro}, {Siopis}, {Smith}, {Sozzetti}, {S{\"u}veges},
  {Torra}, {van Reeven}, {Abbas}, {Abreu Aramburu}, {Accart}, {Aerts},
  {Altavilla}, {{\'A}lvarez}, {Alvarez}, {Alves}, {Anderson}, {Andrei},
  {Anglada Varela}, {Antiche}, {Antoja}, {Arcay}, {Astraatmadja}, {Bach},
  {Baker}, {Balaguer-N{\'u}{\~n}ez}, {Balm}, {Barache}, {Barata}, {Barbato},
  {Barblan}, {Barklem}, {Barrado}, {Barros}, {Bartholom{\'e} Mu{\~n}oz},
  {Bassilana}, {Becciani}, {Bellazzini}, {Berihuete}, {Bertone}, {Bianchi},
  {Bienaym{\'e}}, {Blanco-Cuaresma}, {Boch}, {Boeche}, {Bombrun}, {Borrachero},
  {Bouquillon}, {Bourda}, {Bragaglia}, {Bramante}, {Breddels}, {Brouillet},
  {Br{\"u}semeister}, {Brugaletta}, {Bucciarelli}, {Burlacu}, {Busonero},
  {Butkevich}, {Buzzi}, {Caffau}, {Cancelliere}, {Cannizzaro}, {Carballo},
  {Carlucci}, {Carrasco}, {Casamiquela}, {Castellani}, {Castro-Ginard},
  {Charlot}, {Chemin}, {Chiavassa}, {Cocozza}, {Costigan}, {Cowell}, {Crifo},
  {Crosta}, {Crowley}, {Cuypers}, {Dafonte}, {Damerdji}, {Dapergolas}, {David},
  {David}, {de Laverny}, {De Luise}, {De March}, {de Martino}, {de Souza}, {de
  Torres}, {Debosscher}, {del Pozo}, {Delbo}, {Delgado}, {Delgado}, {Diakite},
  {Diener}, {Distefano}, {Dolding}, {Drazinos}, {Dur{\'a}n}, {Edvardsson},
  {Enke}, {Eriksson}, {Esquej}, {Eynard Bontemps}, {Fabre}, {Fabrizio},
  {Faigler}, {Falc{\~a}o}, {Farr{\`a}s Casas}, {Federici}, {Fedorets},
  {Fernique}, {Figueras}, {Filippi}, {Findeisen}, {Fonti}, {Fraile}, {Fraser},
  {Fr{\'e}zouls}, {Gai}, {Galleti}, {Garabato}, {Garc{\'\i}a-Sedano},
  {Garofalo}, {Garralda}, {Gavel}, {Gavras}, {Gerssen}, {Geyer}, {Giacobbe},
  {Gilmore}, {Girona}, {Giuffrida}, {Glass}, {Gomes}, {Granvik}, {Gueguen},
  {Guerrier}, {Guiraud}, {Guti{\'e}}, {Haigron}, {Hatzidimitriou}, {Hauser},
  {Haywood}, {Heiter}, {Helmi}, {Heu}, {Hilger}, {Hobbs}, {Hofmann}, {Holland},
  {Huckle}, {Hypki}, {Icardi}, {Jan{\ss}en}, {Jevardat de Fombelle}, {Jonker},
  {Juh{\'a}sz}, {Julbe}, {Karampelas}, {Kewley}, {Klar}, {Kochoska}, {Kohley},
  {Kolenberg}, {Kontizas}, {Kontizas}, {Koposov}, {Kordopatis},
  {Kostrzewa-Rutkowska}, {Koubsky}, {Lambert}, {Lanza}, {Lasne}, {Lavigne}, {Le
  Fustec}, {Le Poncin-Lafitte}, {Lebreton}, {Leccia}, {Leclerc},
  {Lecoeur-Taibi}, {Lenhardt}, {Leroux}, {Liao}, {Licata}, {Lindstr{\o}m},
  {Lister}, {Livanou}, {Lobel}, {L{\'o}pez}, {Managau}, {Mann}, {Mantelet},
  {Marchal}, {Marchant}, {Marconi}, {Marinoni}, {Marschalk{\'o}}, {Marshall},
  {Martino}, {Marton}, {Mary}, {Massari}, {Matijevi{\v{c}}}, {Mazeh},
  {McMillan}, {Messina}, {Michalik}, {Millar}, {Molina}, {Molinaro},
  {Moln{\'a}r}, {Montegriffo}, {Mor}, {Morbidelli}, {Morel}, {Morris},
  {Mulone}, {Muraveva}, {Musella}, {Nelemans}, {Nicastro}, {Noval},
  {O'Mullane}, {Ord{\'e}novic}, {Ord{\'o}{\~n}ez-Blanco}, {Osborne}, {Pagani},
  {Pagano}, {Pailler}, {Palacin}, {Palaversa}, {Panahi}, {Pawlak},
  {Piersimoni}, {Pineau}, {Plachy}, {Plum}, {Poggio}, {Poujoulet},
  {Pr{\v{s}}a}, {Pulone}, {Racero}, {Ragaini}, {Rambaux}, {Ramos-Lerate},
  {Regibo}, {Reyl{\'e}}, {Riclet}, {Ripepi}, {Riva}, {Rivard}, {Rixon},
  {Roegiers}, {Roelens}, {Romero-G{\'o}mez}, {Rowell}, {Royer}, {Ruiz-Dern},
  {Sadowski}, {Sagrist{\`a} Sell{\'e}s}, {Sahlmann}, {Salgado}, {Salguero},
  {Sanna}, {Santana-Ros}, {Sarasso}, {Savietto}, {Schultheis}, {Sciacca},
  {Segol}, {Segovia}, {S{\'e}gransan}, {Shih}, {Siltala}, {Silva}, {Smart},
  {Smith}, {Solano}, {Solitro}, {Sordo}, {Soria Nieto}, {Souchay}, {Spagna},
  {Spoto}, {Stampa}, {Steele}, {Steidelm{\"u}ller}, {Stephenson}, {Stoev},
  {Suess}, {Surdej}, {Szabados}, {Szegedi-Elek}, {Tapiador}, {Taris}, {Tauran},
  {Taylor}, {Teixeira}, {Terrett}, {Teyssandier}, {Thuillot}, {Titarenko},
  {Torra Clotet}, {Turon}, {Ulla}, {Utrilla}, {Uzzi}, {Vaillant}, {Valentini},
  {Valette}, {van Elteren}, {Van Hemelryck}, {Vaschetto}, {Vecchiato},
  {Veljanoski}, {Viala}, {Vicente}, {Vogt}, {von Essen}, {Voss}, {Votruba},
  {Voutsinas}, {Walmsley}, {Weiler}, {Wertz}, {Wevers}, {Wyrzykowski},
  {Yoldas}, {{\v{Z}}erjal}, {Ziaeepour}, {Zorec}, {Zschocke}, {Zucker},
  {Zurbach}, \& {Zwitter}}]{Gaia_Collaboration18}
{Gaia Collaboration}, {Babusiaux}, C., {van Leeuwen}, F., {et~al.} 2018, \aap,
  616, A10

\bibitem[{{Green} {et~al.}(2019){Green}, {Schlafly}, {Zucker}, {Speagle}, \&
  {Finkbeiner}}]{Green2019}
{Green}, G.~M., {Schlafly}, E., {Zucker}, C., {Speagle}, J.~S., \&
  {Finkbeiner}, D. 2019, \apj, 887, 93

\bibitem[{{Hayden} {et~al.}(2020){Hayden}, {Bland-Hawthorn}, {Sharma},
  {Freeman}, {Kos}, {Buder}, {Anguiano}, {Asplund}, {Chen}, {De Silva},
  {Khanna}, {Lin}, {Horner}, {Martell}, {Ting}, {Wyse}, {Zucker}, \&
  {Zwitter}}]{Hayden20}
{Hayden}, M.~R., {Bland-Hawthorn}, J., {Sharma}, S., {et~al.} 2020, \mnras,
  493, 2952

\bibitem[{{Horne} \& {Baliunas}(1986)}]{Horne86}
{Horne}, J.~H. \& {Baliunas}, S.~L. 1986, \apj, 302, 757

\bibitem[{{Lamm} {et~al.}(2004){Lamm}, {Bailer-Jones}, {Mundt}, {Herbst}, \&
  {Scholz}}]{Lamm04}
{Lamm}, M.~H., {Bailer-Jones}, C.~A.~L., {Mundt}, R., {Herbst}, W., \&
  {Scholz}, A. 2004, \aap, 417, 557

\bibitem[{{Lebreton} \& {Montalb{\'a}n}(2009)}]{Lebreton09}
{Lebreton}, Y. \& {Montalb{\'a}n}, J. 2009, in The Ages of Stars, ed. E.~E.
  {Mamajek}, D.~R. {Soderblom}, \& R.~F.~G. {Wyse}, Vol. 258, 419--430

\bibitem[{{Lindegren} {et~al.}(2018){Lindegren}, {Hern{\'a}ndez}, {Bombrun},
  {Klioner}, {Bastian}, {Ramos-Lerate}, {de Torres}, {Steidelm{\"u}ller},
  {Stephenson}, {Hobbs}, {Lammers}, {Biermann}, {Geyer}, {Hilger}, {Michalik},
  {Stampa}, {McMillan}, {Casta{\~n}eda}, {Clotet}, {Comoretto}, {Davidson},
  {Fabricius}, {Gracia}, {Hambly}, {Hutton}, {Mora}, {Portell}, {van Leeuwen},
  {Abbas}, {Abreu}, {Altmann}, {Andrei}, {Anglada}, {Balaguer-N{\'u}{\~n}ez},
  {Barache}, {Becciani}, {Bertone}, {Bianchi}, {Bouquillon}, {Bourda},
  {Br{\"u}semeister}, {Bucciarelli}, {Busonero}, {Buzzi}, {Cancelliere},
  {Carlucci}, {Charlot}, {Cheek}, {Crosta}, {Crowley}, {de Bruijne}, {de
  Felice}, {Drimmel}, {Esquej}, {Fienga}, {Fraile}, {Gai}, {Garralda},
  {Gonz{\'a}lez-Vidal}, {Guerra}, {Hauser}, {Hofmann}, {Holl}, {Jordan},
  {Lattanzi}, {Lenhardt}, {Liao}, {Licata}, {Lister}, {L{\"o}ffler},
  {Marchant}, {Martin-Fleitas}, {Messineo}, {Mignard}, {Morbidelli}, {Poggio},
  {Riva}, {Rowell}, {Salguero}, {Sarasso}, {Sciacca}, {Siddiqui}, {Smart},
  {Spagna}, {Steele}, {Taris}, {Torra}, {van Elteren}, {van Reeven}, \&
  {Vecchiato}}]{Lindegren18}
{Lindegren}, L., {Hern{\'a}ndez}, J., {Bombrun}, A., {et~al.} 2018, \aap, 616,
  A2

\bibitem[{{Maldonado} {et~al.}(2015){Maldonado}, {Eiroa}, {Villaver},
  {Montesinos}, \& {Mora}}]{Maldonado15}
{Maldonado}, J., {Eiroa}, C., {Villaver}, E., {Montesinos}, B., \& {Mora}, A.
  2015, \aap, 579, A20

\bibitem[{{Mamajek} \& {Hillenbrand}(2008)}]{Mamajek08}
{Mamajek}, E.~E. \& {Hillenbrand}, L.~A. 2008, \apj, 687, 1264

\bibitem[{{Marshall} {et~al.}(2006){Marshall}, {Robin}, {Reyl{\'e}},
  {Schultheis}, \& {Picaud}}]{Marshall2006}
{Marshall}, D.~J., {Robin}, A.~C., {Reyl{\'e}}, C., {Schultheis}, M., \&
  {Picaud}, S. 2006, \aap, 453, 635

\bibitem[{{McQuillan} {et~al.}(2013){McQuillan}, {Aigrain}, \&
  {Mazeh}}]{Mcquillan13}
{McQuillan}, A., {Aigrain}, S., \& {Mazeh}, T. 2013, \mnras, 432, 1203

\bibitem[{{Messina}(2019)}]{Messina19}
{Messina}, S. 2019, \aap, 627, A97

\bibitem[{{Messina}(2021)}]{Messina21}
{Messina}, S. 2021, \aap, 645, A144

\bibitem[{{Messina} {et~al.}(2010){Messina}, {Desidera}, {Turatto},
  {Lanzafame}, \& {Guinan}}]{Messina10}
{Messina}, S., {Desidera}, S., {Turatto}, M., {Lanzafame}, A.~C., \& {Guinan},
  E.~F. 2010, \aap, 520, A15

\bibitem[{{Messina} {et~al.}(2016){Messina}, {Lanzafame}, {Feiden}, {Millward},
  {Desidera}, {Buccino}, {Curtis}, {Jofr{\'e}}, {Kehusmaa}, {Medhi}, {Monard},
  \& {Petrucci}}]{Messina16}
{Messina}, S., {Lanzafame}, A.~C., {Feiden}, G.~A., {et~al.} 2016, \aap, 596,
  A29

\bibitem[{{Messina} {et~al.}(2017){Messina}, {Millward}, {Buccino}, {Zhang},
  {Medhi}, {Jofr{\'e}}, {Petrucci}, {Pi}, {Hambsch}, {Kehusmaa}, {Harlingten},
  {Artemenko}, {Curtis}, {Hentunen}, {Malo}, {Mauas}, {Monard}, {Muro Serrano},
  {Naves}, {Santallo}, {Savuskin}, \& {Tan}}]{Messina17}
{Messina}, S., {Millward}, M., {Buccino}, A., {et~al.} 2017, \aap, 600, A83

\bibitem[{{Morrell} \& {Naylor}(2019)}]{Morrell19}
{Morrell}, S. \& {Naylor}, T. 2019, \mnras, 489, 2615

\bibitem[{{Nardiello}(2020)}]{Nardiello2020b}
{Nardiello}, D. 2020, \mnras, 498, 5972

\bibitem[{{Nardiello} {et~al.}(2015){Nardiello}, {Bedin}, {Nascimbeni},
  {Libralato}, {Cunial}, {Piotto}, {Bellini}, {Borsato}, {Brogaard}, {Granata},
  {Malavolta}, {Marino}, {Milone}, {Ochner}, {Ortolani}, {Tomasella},
  {Clemens}, \& {Salaris}}]{Nardiello2015}
{Nardiello}, D., {Bedin}, L.~R., {Nascimbeni}, V., {et~al.} 2015, \mnras, 447,
  3536

\bibitem[{{Nardiello} {et~al.}(2019){Nardiello}, {Borsato}, {Piotto},
  {Colombo}, {Manthopoulou}, {Bedin}, {Granata}, {Lacedelli}, {Libralato},
  {Malavolta}, {Montalto}, \& {Nascimbeni}}]{Nardiello2019}
{Nardiello}, D., {Borsato}, L., {Piotto}, G., {et~al.} 2019, \mnras, 490, 3806

\bibitem[{{Nardiello} {et~al.}(2021){Nardiello}, {Deleuil}, {Mantovan},
  {Malavolta}, {Lacedelli}, {Libralato}, {Bedin}, {Borsato}, {Granata}, \&
  {Piotto}}]{2021arXiv210509952N}
{Nardiello}, D., {Deleuil}, M., {Mantovan}, G., {et~al.} 2021, \mnras, 505,
  3767

\bibitem[{{Nardiello} {et~al.}(2016){Nardiello}, {Libralato}, {Bedin},
  {Piotto}, {Ochner}, {Cunial}, {Borsato}, \& {Granata}}]{Nardiello2016}
{Nardiello}, D., {Libralato}, M., {Bedin}, L.~R., {et~al.} 2016, \mnras, 455,
  2337

\bibitem[{{Nardiello} {et~al.}(2020){Nardiello}, {Piotto}, {Deleuil},
  {Malavolta}, {Montalto}, {Bedin}, {Borsato}, {Granata}, {Libralato}, \&
  {Manthopoulou}}]{Nardiello2020a}
{Nardiello}, D., {Piotto}, G., {Deleuil}, M., {et~al.} 2020, \mnras, 495, 4924

\bibitem[{{Oh} {et~al.}(2017){Oh}, {Price-Whelan}, {Hogg}, {Morton}, \&
  {Spergel}}]{Oh17}
{Oh}, S., {Price-Whelan}, A.~M., {Hogg}, D.~W., {Morton}, T.~D., \& {Spergel},
  D.~N. 2017, \aj, 153, 257

\bibitem[{{Pecaut} \& {Mamajek}(2013)}]{Pecaut13}
{Pecaut}, M.~J. \& {Mamajek}, E.~E. 2013, \apjs, 208, 9

\bibitem[{{Pollacco} {et~al.}(2006){Pollacco}, {Skillen}, {Collier Cameron},
  {Christian}, {Irwin}, {Lister}, {Street}, {West}, {Clarkson}, {Evans},
  {Fitzsimmons}, {Haswell}, {Hellier}, {Hodgkin}, {Horne}, {Jones}, {Kane},
  {Keenan}, {Norton}, {Osborne}, {Ryans}, \& {Wheatley}}]{Pollacco06}
{Pollacco}, D., {Skillen}, I., {Collier Cameron}, A., {et~al.} 2006, \apss,
  304, 253

\bibitem[{{Pont} \& {Eyer}(2004)}]{Pont04}
{Pont}, F. \& {Eyer}, L. 2004, \mnras, 351, 487

\bibitem[{{Rebull} {et~al.}(2016){Rebull}, {Stauffer}, {Bouvier}, {Cody},
  {Hillenbrand}, {Soderblom}, {Valenti}, {Barrado}, {Bouy}, {Ciardi},
  {Pinsonneault}, {Stassun}, {Micela}, {Aigrain}, {Vrba}, {Somers}, {Gillen},
  \& {Collier Cameron}}]{Rebull16}
{Rebull}, L.~M., {Stauffer}, J.~R., {Bouvier}, J., {et~al.} 2016, \aj, 152, 114

\bibitem[{{Rebull} {et~al.}(2017){Rebull}, {Stauffer}, {Hillenbrand}, {Cody},
  {Bouvier}, {Soderblom}, {Pinsonneault}, \& {Hebb}}]{Rebull17}
{Rebull}, L.~M., {Stauffer}, J.~R., {Hillenbrand}, L.~A., {et~al.} 2017, \apj,
  839, 92

\bibitem[{{Roberts} {et~al.}(1987){Roberts}, {Lehar}, \& {Dreher}}]{Roberts87}
{Roberts}, D.~H., {Lehar}, J., \& {Dreher}, J.~W. 1987, \aj, 93, 968

\bibitem[{{Soderblom}(2010)}]{Soderblom10}
{Soderblom}, D.~R. 2010, \araa, 48, 581

\bibitem[{{Stauffer} {et~al.}(1998){Stauffer}, {Schultz}, \&
  {Kirkpatrick}}]{Stauffer98}
{Stauffer}, J.~R., {Schultz}, G., \& {Kirkpatrick}, J.~D. 1998, \apjl, 499,
  L199

\bibitem[{{Tang} {et~al.}(2018){Tang}, {Chen}, {Chiang}, {Jose}, {Herczeg}, \&
  {Goldman}}]{Tang18}
{Tang}, S.-Y., {Chen}, W.~P., {Chiang}, P.~S., {et~al.} 2018, \apj, 862, 106

\bibitem[{{Tang} {et~al.}(2019){Tang}, {Pang}, {Yuan}, {Chen}, {Hong},
  {Goldman}, {Just}, {Shukirgaliyev}, \& {Lin}}]{Tang19}
{Tang}, S.-Y., {Pang}, X., {Yuan}, Z., {et~al.} 2019, \apj, 877, 12

\bibitem[{{Tokovinin} \& {Brice{\~n}o}(2018)}]{Tokovinin18}
{Tokovinin}, A. \& {Brice{\~n}o}, C. 2018, \aj, 156, 138

\bibitem[{{Zechmeister} \& {K{\"u}rster}(2009)}]{Zechmeister09}
{Zechmeister}, M. \& {K{\"u}rster}, M. 2009, \aap, 496, 577

\bibitem[{{Zhang} {et~al.}(2019){Zhang}, {Zhao}, {Oswalt}, {Fang}, {Zhao},
  {Liang}, {Ye}, \& {Zhong}}]{Zhang19}
{Zhang}, J., {Zhao}, J., {Oswalt}, T.~D., {et~al.} 2019, \apj, 887, 84

\end{thebibliography}

\begin{appendix}
\section{Table}
In the following Table\,\ref{tab-period}, the Group X members with ID number, G magnitude, de-reddened colour, rotation period and uncertainty, grade of confidence, and TESS sector of observations are listed.
\onecolumn
\begin{center}
\begin{longtable}{l@{\hspace{.1cm}}l@{\hspace{.1cm}}c@{\hspace{.5cm}}c@{\hspace{.5cm}}c@{\hspace{.1cm}}c@{\hspace{.5cm}}l@{\hspace{.5cm}}}
\caption{Rotation periods of Group X candidate members.}\label{tab-period}\\%
\hline\hline 
Sequ. \# & TIC number         & G     &  (G$-$K)$_0$   & P $\pm$ $\sigma$&  confidence   & TESS Sector  \\%
            &                 &          (mag)     &   (mag)        &  (d)   &   grade   &         \\%
\hline
\endfirsthead
\caption{continued.}\\%
\hline\hline
     Sequ. \# & TIC number            & G     &  (G$-$K)$_0$    & P $\pm$ $\sigma$&  confidence   & TESS Sector \\%
            &                 &   (mag)               &   (mag)      &  (d)   &    grade  &         \\%
\hline
\hline

\endhead

\hline
\endfoot

\hline 
\endlastfoot

{5}   &        0334518873   &   16.025   &    3.962   &   6.9  $\pm$    1.0   &       B   &                          15   \\
  6   &        0136951754   &   14.773   &    3.675   &   0.98  $\pm$    0.02   &       A   &                    15/16/22   \\
  7   &        0155871409   &   16.881   &    4.006   &   0.217  $\pm$    0.001   &       A   &                       16/22   \\
{8}   &        0155856633   &    8.973   &    1.212   &   5.08  $\pm$    0.16   &       A   &                    15/16/22   \\
  9   &        0142386740   &   15.576   &    3.869   &   0.788  $\pm$    0.012   &       A   &                       21/22   \\
 11   &        0417937937   &   16.224   &    3.879   &   1.042  $\pm$    0.023   &       A   &                          15   \\
 12$^{\star}$    &        0159159752   &   16.166   &    3.841   &   1.22  $\pm$    0.03   &       A   &                    15/16/22   \\
                &                       &           &           &    6.5 $\pm$ 0.8          &           A &                16/22 \\
 13   &        0142413357   &   16.019   &    3.905   &   1.01  $\pm$    0.02   &       A   &                 15/16/21/22   \\
 15$^{\star}$    &        0165454079   &   14.882   &    3.676   &   1.30  $\pm$    0.03   &       A   &                    15/16/22   \\
 16   &        0459220753   &   14.843   &    3.751   &   1.92  $\pm$    0.07   &       A   &                    15/16/22   \\
 17$^{\star a}$    &        0524500866   &   16.566   &    3.723   &   0.517  $\pm$    0.005   &       A   &                          15   \\
                &                       &           &               & 0.405 $\pm$ 0.004      &       A   &           \\
 18   &        0459220751   &   15.652   &    3.957   &   1.10  $\pm$    0.02   &       A   &                    15/16/22   \\
 20   &        0459221489   &   13.006   &    3.012   &   6.9  $\pm$    1.0   &       B   &                       15/22   \\
 21   &        0459221499   &    7.513   &    0.653   &   0.373  $\pm$    0.003   &       A   &                    15/16/22   \\
 22   &        0165464384   &   14.846   &    3.609   &   4.09  $\pm$    0.35   &       A   &                    15/16/22   \\
 23   &        0446174335   &   15.170   &    3.725   &   3.60  $\pm$    0.28   &       A   &                    16/22/23   \\
 26   &        0158460920   &   16.041   &    3.845   &   0.81  $\pm$    0.01   &       A   &                          15   \\
 27$^{\star}$    &        0288454252   &   15.519   &    3.930   &   1.17  $\pm$    0.03   &       A   &                    16/22/23   \\
                &                       &           &           &       0.587 $\pm$ 0.004 &   A           &                               \\
 29$^{\star a}$    &        1001374231   &   15.710   &    3.602   &   1.97  $\pm$    0.08   &       A   &                    15/16/22   \\
 30$^{\star a}$    &        1001374230   &   15.444   &    3.512   &   0.487  $\pm$    0.005   &       A   &                    15/16/22   \\
{31}   &        0332277847   &   15.307   &    3.714   &   6.13  $\pm$    0.78   &       A   &                       15/16   \\
 32   &        0141814573   &   15.578   &    3.650   &   1.40  $\pm$    0.04   &       A   &                       15/22   \\
 33   &        0311068695   &   16.206   &    3.699   &   0.633  $\pm$    0.008   &       A   &                 15/16/21/22   \\
 35   &        0311068638   &   11.854   &    2.382   &   8.9  $\pm$    1.7   &       B   &                          16   \\
 37    &        0288512352   &   15.273   &    3.977   &   0.496  $\pm$    0.005   &       A   &                    16/22/23   \\
 38   &        0332313010   &   14.343   &    3.610   &   4.23  $\pm$    0.37   &       A   &                 15/16/22/23   \\
 39   &        0332312964   &    9.580   &    1.341   &   4.98  $\pm$    0.52   &       A   &                 15/16/22/23   \\
      &                     &            &            &   0.363  $\pm$    0.004  &       A   &               15/16/22/23 \\
 40   &        0310995545   &   14.387   &    3.685   &   0.481  $\pm$    0.005   &       A   &                       16/22   \\
 41   &        0310996926   &   15.158   &    3.682   &   1.23  $\pm$    0.03   &       A   &                 15/16/22/23   \\
 42    &        0141819826   &    9.534   &    1.337   &   7.1  $\pm$    1.2   &       B   &                 16/15/22/23   \\
 43   &        0141819348   &   15.914   &    3.976   &   0.453  $\pm$    0.004   &       A   &                          15   \\
 45   &        0311001628   &    6.009   &    0.010   &   0.602  $\pm$    0.008   &       A   &                          16   \\
 46   &        0311001756   &   14.018   &    3.494   &   0.75  $\pm$    0.02   &       A   &                 15/16/22/23   \\
 47   &        0311002115   &   15.395   &    3.834   &   1.01  $\pm$    0.02   &       A   &                       15/16   \\
 48   &        0165407465   &   13.281   &    3.249   &   0.79  $\pm$    0.01   &       A   &                    16/22/23   \\
 49   &        0310338842   &   10.451   &    1.972   &   7.8  $\pm$    1.2   &       A   &                 15/16/22/23   \\
 50   &  0219032664  &  14.129  &   3.323   &  7.2 $\pm$ 1.1   &       B   & 16/23   \\
 51   &        0141861147   &    5.655   &   -0.121   &   1.38  $\pm$    0.04   &       A   &                 15/16/22/23   \\
 52   &        0141862036   &   13.944   &    3.510   &   0.560  $\pm$    0.006   &       A   &                    15/16/22   \\
 53   &        0219034610   &   16.524   &    4.075   &   0.1942  $\pm$    0.0008   &       A   &                       15/16   \\
 54   &        0141863294   &   10.808   &    1.937   &   8.1  $\pm$    1.8   &       B   &                       16/22   \\
 55   &        0219034788   &   15.352   &    3.788   &   1.85  $\pm$    0.07   &       A   &                 15/16/22/23   \\
 56   &        0141863170   &   16.361   &    3.943   &   1.51  $\pm$    0.04   &       B   &                          22   \\
 57   &        0219036430   &   14.063   &    3.388   &   2.87  $\pm$    0.18   &       A   &                    16/22/23   \\
 58   &        0141915639   &   14.343   &    3.536   &   0.369  $\pm$    0.002   &       A   &                    15/16/22   \\
 60   &        0219041670   &   14.105   &    3.455   &   5.61  $\pm$    0.69   &       A   &                       16/22   \\
 61   &        0467178971   &   16.465   &    3.956   &   1.87  $\pm$    0.07   &       A   &                 15/16/22/23   \\
 62   &        0219065608   &   15.942   &    3.898   &   0.86  $\pm$    0.016   &       A   &                    16/22/23   \\
 63  &        0441640476   &  16.5047   &    3.859  &  7.0   $\pm$   1.0    &      A  &                  22/16 \\
      &                     &            &            &  0.464 $\pm$   0.003  &      A  &                   22/16 \\
 64   &        0332313458   &   16.022   &    3.903   &   0.566  $\pm$    0.006   &       A   &                       22/23   \\
 67   &        0154256770   &   15.771   &    3.776   &   1.04  $\pm$    0.02   &       A   &                       16/22   \\
 69   &        0198147567   &   13.259   &    3.121   &   6.49  $\pm$    0.93   &       A   &                       16/23   \\
 70   &        0198147621   &   16.828   &    4.043   &   0.79  $\pm$    0.01   &       A   &                       16/22   \\
 71   &        0233437236   &   10.223   &    1.638   &   6.9  $\pm$    1.0   &       A   &                 15/16/22/23   \\
 72   &        0233458510   &   14.707   &    3.660   &   6.44  $\pm$    0.88   &       A   &                          16   \\
 73   &        0441661202   &   16.729   &    4.028   &   0.389  $\pm$    0.003   &       A   &                          15   \\
 74   &        0198154161   &   11.066   &    2.145   &   9.3  $\pm$    1.9   &       B   &                          16   \\
 75   &        0023871511   &   13.510   &    3.215   &   4.45  $\pm$    0.43   &       B   &                          16   \\
 76   &        0233462658   &   12.032   &    2.552   &   9.5  $\pm$    1.2   &       A   &                       22/15   \\
 77   &        0010728867   &   11.047   &    1.923   &   7.6  $\pm$    1.2   &       A   &                    16/22/23   \\
 78   &        0441687813   &   13.891   &    3.365   &   5.83  $\pm$    0.70   &       A   &                          15   \\
 79   &        0332349583   &   15.789   &    3.889   &   5.80  $\pm$    0.67   &       A   &                          22   \\
 80   &        0166053959   &    8.185   &    0.867   &   0.88  $\pm$    0.02   &       A   &                       15/16   \\
 82   &        0168699315   &   15.759   &    3.832   &   0.530  $\pm$    0.006   &       A   &                    16/22/23   \\
 83   &        0309721363   &   15.515   &    3.806   &   1.89  $\pm$    0.07   &       A   &                       22/23   \\
 84   &        0309751966   &   14.992   &    3.630   &   1.57  $\pm$    0.05   &       A   &                    16/22/23   \\
 85   &        0441694341   &   17.133   &    4.030   &   0.442  $\pm$    0.004   &       A   &                       16/23   \\
 86   &        0166065826   &   15.925   &    3.910   &   0.665  $\pm$    0.009   &       A   &                          15   \\
 88   &        0310379752   &    6.544   &    0.166   &   0.561  $\pm$    0.487   &       A   &                          22   \\
 89   &        0168708111   &   14.614   &    3.845   &   1.23  $\pm$    0.03   &       A   &                          16   \\
 90   &        0441697195   &   15.440   &    3.752   &   3.01  $\pm$    0.18   &       A   &                       15/16   \\
 91$^{\star a}$   &        1001276338   &    9.975   &    1.576   &   6.21  $\pm$    0.80   &       A   &                 15/16/22/23   \\
 93   &        0445859773   &    8.425   &    0.947   &   1.28  $\pm$    0.03   &       A   &                       15/16   \\
 94   &        0310394393   &   15.011   &    3.643   &   1.36  $\pm$    0.03   &       A   &                       22/23   \\
 95   &        0313322899   &    6.886   &    0.307   &   0.546  $\pm$    0.006   &       A   &                          22   \\
 96   &        0154382743   &   13.754   &    3.247   &   7.6  $\pm$    1.1   &       A   &                          22   \\
 97   &        0154382045   &   14.498   &    3.508   &   4.67  $\pm$    0.43   &       A   &                 22/23/15/16   \\
 98   &        0441701176   &   15.685   &    3.824   &   0.74  $\pm$    0.01   &       A   &                 15/16/22/23   \\
99   &        0445860782   &   14.968   &    3.656   &   9.9  $\pm$    2.0   &       A   &                       15/16   \\
100   &        0441702640   &   13.784   &    3.276   &   0.82  $\pm$    0.01   &       A   &                       22/23   \\
101   &        0441703294   &   12.340   &    2.661   &   9.00  $\pm$    0.40   &       A   &                          22   \\
102   &        0166089535   &    7.131   &    1.057   &   3.14  $\pm$    0.05   &       A   &                 15/16/21/22   \\
{103}   &        0441704140   &   14.348   &    3.729   &   5.47  $\pm$    0.66   &       A   &                          16   \\
104   &        0441705536   &   16.279   &    4.090   &   0.67  $\pm$    0.01   &       A   &                       16/23   \\
105   &        0459246945   &    9.524   &    1.283   &   4.27  $\pm$    0.38   &       A   &                 15/16/22/23   \\
107   &        0441710042   &   13.671   &    3.247   &   3.77  $\pm$    0.28   &       B   &                          23   \\
{108$^{\star}$}   &        0441709852   &   16.784   &    4.188   &   3.78  $\pm$    0.31   &       A   &                       16/23   \\
109   &        0441711425   &   15.665   &    3.784   &   1.62  $\pm$    0.05   &       A   &                       16/23   \\
112   &        0441711658   &   15.931   &    3.902   &   2.20  $\pm$    0.10   &       A   &                    16/22/23   \\
115   &        0166177052   &   15.628   &    3.908   &   1.56  $\pm$    0.05   &       A   &                 15/16/22/23   \\
{116}   &        0313338124   &   12.485   &    2.984   &   2.73  $\pm$    0.15   &       A   &                    15/16/23   \\
117   &        0166179430   &   14.660   &    3.519   &   1.66  $\pm$    0.05   &       A   &                 15/16/22/23   \\
119   &        0166180049   &   15.623   &    3.788   &   1.47  $\pm$    0.04   &       A   &                    15/16/21   \\
122   &        0298162216   &   12.025   &    2.428   &   9.11  $\pm$    0.58   &       A   &                 16/15/22/23   \\
123   &        0232980303   &   15.685   &    3.764   &   0.88  $\pm$    0.01   &       A   &                 15/16/22/23   \\
124   &        0298163080   &    9.518   &    1.244   &   4.93  $\pm$    0.50   &       A   &                    15/22/23   \\
125   &        0255807075   &   16.191   &    3.953   &   0.489  $\pm$    0.005   &       A   &                    16/22/23   \\
126   &        0159613447   &   10.267   &    1.518   &   6.29  $\pm$    0.82   &       A   &              15/16/21/22/23   \\
127   &        0158462948   &    9.685   &    1.427   &   4.98  $\pm$    0.51   &       A   &                 15/16/22/23   \\
129   &        0158496328   &   16.504   &    3.963   &   0.79  $\pm$    0.01   &       A   &                    16/15/22   \\
130   &        0159628504   &   15.767   &    3.876   &   2.40  $\pm$    0.12   &       A   &                 15/16/22/23   \\
133   &        0159631183   &    8.027   &    0.83   &   0.89  $\pm$    0.01   &       A   &                 15/16/22/23   \\
134   &        0155899586   &   12.353   &    2.675   &  10.0  $\pm$    2.0   &       B   &                          23   \\
{135}   &        0159636302   &   12.123   &    2.598   &   1.09  $\pm$    0.02   &       A   &                 15/16/22/23   \\
136   &        0316420351   &   16.180   &    3.863   &   0.661  $\pm$    0.009   &       A   &                 15/16/22/23   \\
137$^{\star}$   &        0316450376   &   16.229   &    3.847   &   0.40  $\pm$    0.02   &       A   &                    15/16/21   \\
            &                       &               &           &   0.331 $\pm$    0.003  &       A    &                               \\
138   &        0154357603   &   15.785   &    3.923   &   0.438  $\pm$    0.004   &       A   &                    16/22/23   \\
139   &        0158541117   &    9.807   &    1.688   &   6.65  $\pm$    0.88   &       A   &                       22/15   \\
{140}   &        0158563246   &   17.750   &    4.359   &   7.7  $\pm$    1.2   &       A   &                    15/16/22   \\
141$^{\star}$   &        0282920711   &   16.630   &    3.944   &   0.695  $\pm$    0.009   &       A   &              14/15/16/21/23   \\
                &                       &           &           &       7.0 $\pm$ 1.0           &       A                   &       \\
143$^{\star}$   &        0161024760   &   10.089   &    1.609   &   6.48  $\pm$    0.83   &       A   &                    22/16/23   \\
144   &        0310003595   &   12.626   &    2.761   &   7.7  $\pm$    1.2   &       A   &                 15/16/22/23   \\
145   &        0158579468   &   11.521   &    2.126   &   7.9  $\pm$    1.3   &       B   &                          15   \\
146   &        0161029191   &    9.817   &    1.355   &   5.34  $\pm$    0.63   &       A   &                    16/22/23   \\
147   &        0158617635   &   16.223   &    3.823   &   0.570  $\pm$    0.006   &       A   &                 15/16/22/23   \\
{148}$^{\star}$   &        1102311836   &    9.826   &    1.082   &   0.663  $\pm$    0.006   &       A   &                 15/16/22/23   \\
{149}$^{\star}$   &        1102311837   &    8.139   &    1.102   &   4.64  $\pm$    0.45   &       A   &                 15/16/22/23   \\
{151}   &        0462573387   &   11.859   &    2.306   &   3.71  $\pm$    0.28   &       A   &                 15/16/22/23   \\
152   &        0462572935   &   11.742   &    2.222   &   8.8  $\pm$    1.4   &       A   &                          16   \\
153   &        0137832480   &    9.276   &    1.144   &   3.57  $\pm$    0.26   &       A   &                 15/16/22/23   \\
155   &        0137834492   &    7.363   &    0.359   &   0.1050  $\pm$    0.0002   &       A   &                    16/22/23   \\
156   &        0137834173   &   14.564   &    3.396   &   2.87  $\pm$    0.16   &       B   &                          23   \\
157   &        0137834385   &   15.488   &    3.765   &   2.50  $\pm$    0.13   &       A   &                    15/22/23   \\
158   &        0137834559   &    9.502   &    1.499   &   5.37  $\pm$    0.60   &       A   &                 15/16/22/23   \\
159$^{\star a}$   &        0202425640   &   11.511   &    2.211   &   9.23  $\pm$    0.70   &       A   &                 15/16/22/23   \\
160   &        0137842121   &   17.289   &    4.056   &   0.220  $\pm$    0.001   &       A   &                 15/16/22/23   \\
161   &        0137842286   &   16.737   &    3.990   &   0.646  $\pm$    0.008   &       A   &                 15/16/22/23   \\
164   &        0232541198   &   10.368   &    1.448   &   5.21  $\pm$    0.54   &       A   &              14/15/16/21/23   \\
165   &        0165628355   &   12.182   &    2.580   &   9.21  $\pm$    0.80   &       A   &                       15/23   \\
167$^{\star}$   &        1102236385   &   10.939   &    3.941   &   7.2  $\pm$    1.1   &       A   &                 16/22/23/24   \\
169   &        0165651031   &   15.232   &    3.594   &   2.72  $\pm$    0.15   &       A   &                          15   \\
170   &        0165651137   &   14.541   &    3.546   &   5.03  $\pm$    0.52   &       A   &                          15   \\
171   &        0165650305   &   12.390   &    2.435   &   9.4  $\pm$    1.9   &       B   &                       16/23   \\
172   &        0202468203   &   16.885   &    4.052   &   0.93  $\pm$    0.01   &       A   &              14/15/16/21/23   \\
173   &        0165652683   &   14.687   &    3.542   &   1.73  $\pm$    0.06   &       A   &                    16/22/24   \\
174   &        0165652279   &   16.217   &    3.811   &   1.13  $\pm$    0.02   &       A   &                    23/16/24   \\
175   &        0282940237   &   13.584   &    3.043   &   7.4  $\pm$    1.1   &       A   &                    14/15/16   \\
{177}   &        0193969477   &   16.092   &    3.856   &   7.3  $\pm$    1.1   &       A   &                       16/23   \\
179   &        0202490066   &   17.006   &    4.088   &   0.563  $\pm$    0.006   &       B   &                          15   \\
180   &        0202503605   &   15.634   &    3.700   &   0.72  $\pm$    0.01   &       A   &                    15/22/24   \\
181   &        0165715493   &   16.817   &    3.921   &   1.50  $\pm$    0.05   &       A   &                          16   \\
182   &        0165719269   &   10.936   &    1.894   &   7.3  $\pm$    1.2   &       A   &                          16   \\
183$^{\star}$   &        0193991974   &   16.071   &    3.958   &   1.014  $\pm$    0.0923   &       A   &                          24   \\
184   &        0193991973   &   16.088   &    3.956   &   1.125  $\pm$    0.004   &       A   &                          16   \\
187   &        0165735666   &   16.304   &    3.845   &   0.70  $\pm$    0.01   &       A   &                       16/24   \\
188   &        0202510436   &   10.866   &    1.874   &   7.5  $\pm$    1.2   &       A   &                 15/16/22/23   \\
190   &        0165791639   &   16.031   &    3.798   &   1.32  $\pm$    0.03   &       A   &                          16   \\
192   &        0405526687   &   17.727   &    4.120   &   0.235  $\pm$    0.001   &       A   &                          15   \\
193   &        0159769293   &   10.879   &    1.812   &   7.4  $\pm$    1.2 &       A   &                       16/23   \\
196   &        0405580571   &   15.121   &    3.587   &   2.47  $\pm$    0.13   &       A   &                 16/22/23/24   \\
197   &        0159784194   &   17.006   &    3.940   &   1.02  $\pm$    0.02   &       A   &                    16/23/24   \\
200   &        0159838724   &   17.298   &    4.155   &   0.4270  $\pm$    0.004   &       A   &                          16   \\
201   &        0161723105   &   15.028   &    3.653   &   2.05  $\pm$    0.08   &       A   &                    22/23/24   \\
202   &        0161723004   &   16.816   &    4.008   &   0.81  $\pm$    0.01   &       A   &                          22   \\
203   &        0159840416   &   16.822   &    4.041   &   0.546  $\pm$    0.006   &       A   &                    23/16/24   \\
204   &        0286925711   &   12.708   &    2.572   &   9.8  $\pm$    1.9   &       B   &                          23   \\
205   &        0161728466   &   13.431   &    2.923   &   6.49  $\pm$    0.93   &       B   &                       16/24   \\
206   &        0159871552   &   14.833   &    3.450   &   5.40  $\pm$    0.64   &       A   &                    16/23/24   \\
207   &        0159871737   &    9.733   &    1.420   &   4.58  $\pm$    0.46   &       A   &                    16/23/24   \\
208$^{\star}$   &        0159871715   &    8.931   &    1.230   &   0.80  $\pm$    0.047   &       A   &                          24   \\
209   &        0159873822   &   11.325   &    1.871   &   7.5  $\pm$    1.2   &       A   &                       16/23   \\
210   &        0161744704   &   17.810   &    4.252   &   0.226  $\pm$    0.001   &       A   &                       24/16   \\
211   &        0159879031   &    9.774   &    1.222   &   4.28  $\pm$    0.40   &       A   &                    16/23/24   \\
213   &        0159922985   &   10.201   &    1.414   &   5.20  $\pm$    0.54   &       A   &                 23/16/24/25   \\
214   &        0219479795   &   10.432   &    1.682   &   7.0  $\pm$    1.0   &       B   &                          23   \\
217$^{\star a}$   &        0219503289   &   14.821   &    3.320   &   0.777  $\pm$    0.003   &       A   &                       24/25   \\
\hline
\multicolumn{7}{l}{The periods of \#29\&\#30 and to \#148\&\#149 are arbitrarily assigned, these systems }\\
\multicolumn{7}{l}{being unresolved; $^{\star}$ unresolved systems in TESS photometry; $^a$ (G$-$K$_s)_0$ }\\
\multicolumn{7}{l}{colour corrected for binarity.}\\
\end{longtable}
\end{center}

\end{appendix}

\end{document}